\documentclass[reprint, superscriptaddress, amsmath, amssymb, aps, prl]{revtex4-2}

\usepackage{amsmath}
\usepackage{amssymb}
\usepackage{indentfirst}
\usepackage{commath}
\usepackage{graphicx}
\usepackage[caption=false,position=bottom,labelfont={bf}]{subfig}
\usepackage{overpic}
\usepackage{dcolumn}
\usepackage{bm}
\usepackage{booktabs}
\usepackage{dsfont}
\usepackage{setspace}
\usepackage{siunitx}
\usepackage{multirow}

\usepackage{xcolor}
\definecolor{myred}{rgb}{0.8,0.1,0.2}
\definecolor{myblue}{rgb}{0.1,0.2,0.6}

\usepackage[colorlinks,linktocpage,hypertexnames=false]{hyperref}
\hypersetup{
    colorlinks=true,
    linktoc=all,
    linkcolor={myred},
    citecolor={myblue},
    urlcolor={myblue},
}

\usepackage{xparse}
\NewDocumentCommand{\bra}{o m}{
  \IfNoValueTF{#1}
    {\left\langle #2 \right|}
    {\langle #2 |}
}
\NewDocumentCommand{\ket}{o m}{
  \IfNoValueTF{#1}
    {\left| #2 \right\rangle}
    {| #2 \rangle}
}
\ExplSyntaxOn
\NewDocumentCommand{\expt}{O{true} m G{} G{}}{
  \str_if_eq:nnTF {#1} {true}
    {
      \tl_if_empty:nTF {#3}
        {\left\langle #2 \right\rangle}
        {
          \tl_if_empty:nTF {#4}
            {\left\langle #2 \middle| #3 \right\rangle}
            {\left\langle #2 \middle| #3 \middle| #4 \right\rangle}
        }
    }
    {
      \tl_if_empty:nTF {#3}
        {\langle #2 \rangle}
        {
          \tl_if_empty:nTF {#4}
            {\langle #2 | #3 \rangle}
            {\langle #2 | #3 | #4 \rangle}
        }
    }
}
\ExplSyntaxOff


\begin{document}

\title{
Quantum geometric localization length and localization criticality\\
in an ideally flat Chern band
}

\author{Xu-Cheng Wang}
\affiliation{State Key Laboratory of Surface Physics, Fudan University, Shanghai 200433, China}
\affiliation{Center for Field Theory and Particle Physics, Department of Physics, Fudan University, Shanghai 200433, China}

\author{Yang Qi}
\email{qiyang@fudan.edu.cn}
\affiliation{State Key Laboratory of Surface Physics, Fudan University, Shanghai 200433, China}
\affiliation{Center for Field Theory and Particle Physics, Department of Physics, Fudan University, Shanghai 200433, China}
\affiliation{Hefei National Laboratory, Hefei 230088, China}

\date{\today}

\begin{abstract}
We propose that the localization length in an isolated, ideally flat Chern band is set by quantum geometry.
We explore the corresponding localization transition and its critical scaling
by applying transfer matrix calculations in the maximally localized hybrid Wannier basis,
whose spatial spread is exactly characterized by a quantum geometric length.
Remarkably, upon tuning the quantum metric of the Chern band, we observe a crossover
from a universal regime controlled by the Dirac fixed point
to a non-universal regime with continuously varying critical exponents.
Within the universal regime,
the localization length exhibits a pronounced linear dependence on the quantum geometric length,
supporting its quantum geometric nature.
These findings provide a novel quantum geometric perspective on
the localization in quantum Hall systems such as twisted moiré superlattices,
and shed new light on the long-standing controversy
over the criticality of the integer quantum Hall transition.
\end{abstract}

\maketitle

{\color{myblue}\it Introduction.---}
The observation of quantized conductance in Landau levels (LL) and topological Chern insulators
has marked a significant milestone in condensed matter physics.
It is well-established that
a nonzero Chern number gives rise to the quantized conductance,
while disorders play a crucial role in stabilizing the conductance plateaus.
In two dimensions (2D) and in the thermodynamic limit,
disorder localizes all quantum states except those at the critical energy $E_c$,
where the localization length diverges according to~\cite{evers2008anderson}
\begin{equation}
    \xi(x) = \xi_0 \abs{x}^{-\nu},
\end{equation}
with $x=(E-E_c)/E_c$ and $\nu$ the critical exponent.

Conventionally, observing the integer and fractional Hall plateaus in 2D electron gas (2DEG)
requires ultra-high carrier mobility up to $10^5\sim 10^7\ \text{cm}^2/\text{Vs}$~\cite{klitzing1980new,
tsui1982two-dimensional,chung2021ultra-high-quality}.
In contrast, the recently reported zero-field integer and, especially, fractional Hall states
in twisted bilayer $\text{MoTe}_2$~\cite{cai2023signatures,park2023observation,xu2023observation,bernevig2025fractional}
typically appear at much lower mobilities of order $10^3\sim 10^4\ \text{cm}^2/\text{Vs}$.
On one hand, moderate disorder can stabilize and extend the integer plateau;
however, if the disorder becomes excessively strong,
it causes the fractional plateau to be masked by the integer one
and even drives a transition to the topologically trivial state.
The observation that pronounced fractional plateaus persist in twisted moiré materials,
despite mobilities significantly inferior to those in 2DEG,
suggests that the Chern bands experience weaker localization than Landau levels.
Since the plateau width is directly related to the mobility edge $E_m$,
given by $E_m/E_c=1+(\xi_0/L)^{1/\nu}$,
a central question then arises concerning the governing factor that
controls the localization length $\xi_0$ in (fractional) Chern insulators under comparable disorder strength.

\begin{figure}[htbp]
    \centering\hspace{0cm}
    \includegraphics[width=.75\columnwidth]{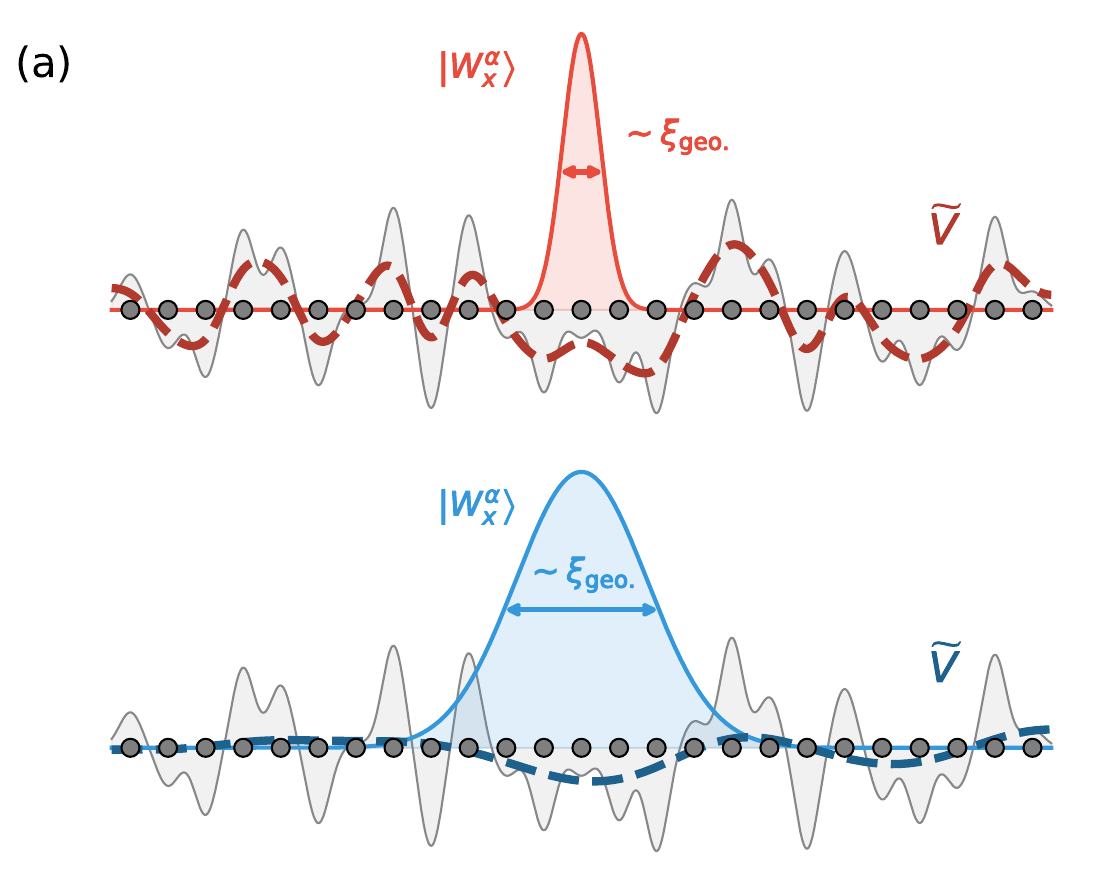}
    \centering\hspace{0cm}\vspace{-.25cm}
    \includegraphics[width=1\columnwidth]{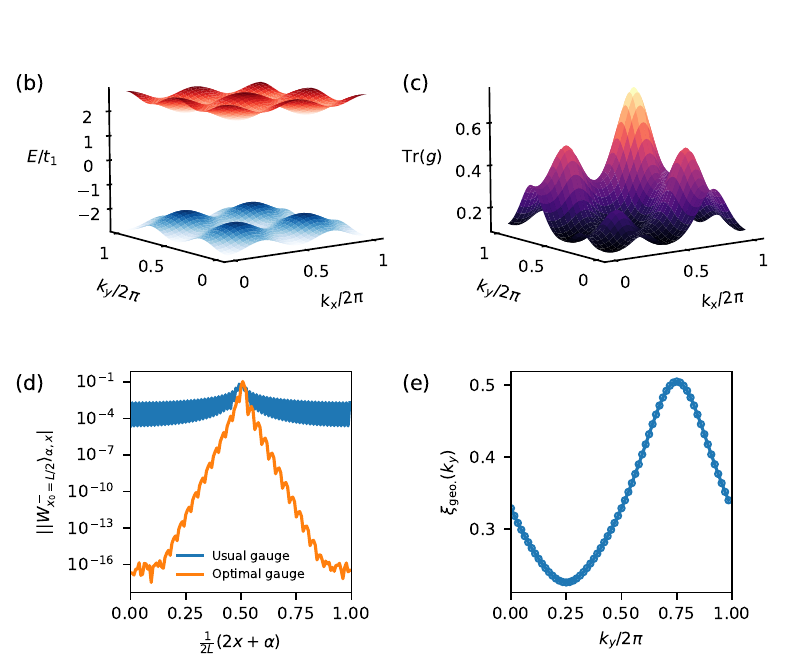}
    \caption{%
        (a) Schematic plot of Eq.~\eqref{eq:argument} that
        ideally flat Chern band with larger quantum geometric length experiences weaker localization.
        We denote the bare disorder potential V as the grey peaks and valleys randomly distributed among the atoms.
        The optimal Wannier spread is a quantum geometric quantity,
        and the effective disorder potential $\widetilde{V}$ can be regarded as an average of bare potentials
        within the wave packet characterized by the Wannier spread.
        (b)-(e) $\pi$-flux model with $\sqrt{2}t_2/t_1=1$.
        (b)(c) The optimally flat band and the corresponding quantum metric of the lower band.
        (d) Maximally localized hybrid Wannier function (orange) as compared to normal Wannier basis (blue) for the lower band.
        We plot $\ket[false]{W^{-}_{k_y=0,x=L/2}}_{\alpha,x,y=0}$ to demonstrate the decay in $x$ direction.
        (e) Quantum geometric length $\xi_\text{geo.}(k_y)$ for the lower band.
    }
    \label{fig:fig1}
\end{figure}

To simplify the problem, we consider an ideally flat Chern band in the presence of local and uncorrelated disorder,
and focus on the weak-disorder regime where the disorder strength is much smaller than the band gap.
Therefore, each Chern band can be considered isolated and we ignore the band mixing.
For such a system, the only intrinsic length scale is a quantum geometric length $\xi_\text{geo.}$,
the counterpart of magnetic length $l_B$ in LLs.
As defined in Eqs.~\eqref{eq:omega-min}\eqref{eq:geometric-length},
$\xi_\text{geo.}$ is related to the quantum metric of Bloch states~\cite{liu2024quantum}.
Our primary claim in this work is that \textit{the localization length $\xi(E)$ in the ideally flat Chern band is a quantum geometric length scale such that}
\begin{equation}\label{eq:argument}
    \xi_0\sim \xi_\text{geo.}.
\end{equation}
Recall that the quantum geometric tensor~\cite{liu2024quantum,cheng2013quantum} for the isolated band
is defined as
\begin{equation}\begin{aligned}
    Q_{\mu\nu}(k)
    &= \expt{\partial_\mu u_k}{\partial_\nu u_k} - \expt{\partial_\mu u_k}{u_k} \expt{u_k}{\partial_\nu u_k}\\
    &\stackrel{!}{=} g_{\mu\nu}-\frac{i}{2}F_{\mu\nu},
\end{aligned}\end{equation}
with $g_{\mu\nu}$ the quantum metric, or Fubini-Study (FS) metric, measuring the invariant distance of Bloch states,
and $F_{\mu\nu}$ the Berry curvature.
It has been recognized~\cite{roy2014band,liu2024quantum} that
the quantum geometry is fundamental in stabilizing the fractional Chern insulator.
Recent studies further showed that a quantum metric length dictates
the spatial extent of topological boundary modes~\cite{ma2025universal,luo2025tunable}.
Here we emphasize that
the spatial spread of the maximally localized hybrid Wannier function of flat Chern band is exactly characterized by
the quantum geometric length $\xi_\text{geo.}$, as illustrated in Fig.~\ref{fig:fig1}(a).
In general, a Wannier wave packet with a large spatial spread
experiences the averaged effective disorder potential $\widetilde{V}$,
where short-wavelength components of $V$ are filtered out.
$\widetilde{V}$ is typically smoother and weaker in amplitude than the bare disorder potential $V$.
As a result, at fixed bare disorder strength,
a Chern band with a large quantum geometric length is effectively protected from localization
due to the softened effective potential $\widetilde{V}$,
and therefore hosts a large localization length $\xi_0$, supporting our argument in Eq.~\eqref{eq:argument}.
Specifically, for LLs in 2DEG, the quantum metric is ideally flat
and its quantum geometric length in Eqs.~\eqref{eq:omega-min}\eqref{eq:geometric-length} simply reduces to the magnetic length $l_B$,
which is known to control the spatial spread of LL wavefunctions.
Therefore, we expect $\xi_0\sim l_B$ for LLs.

To complement this intuitive view rigorously, we establish Eq.~\eqref{eq:argument} by studying the localization transition of
an isolated, ideally flat Chern band with tunable quantum geometric length.
Using transfer matrix calculations novelly in the maximally localized hybrid Wannier basis,
we extract the critical exponent and localization length.
It is found that in the universal regime of the unitary class,
which is the same universality class as the integer quantum Hall transition (IQHT),
the evolution of localization length is faithfully tracked by the quantum geometric length $\xi_\text{geo.}$
and a linear relation between $\xi(E)$ and $\xi_\text{geo.}$ is revealed.
The associated quantum geometric mobility edge is further predicted.
We believe this quantum geometric perspective on the localization length and mobility edge
will advance our understanding of plateau transitions in (fractional) quantum anomalous Hall systems~\cite{chang2023colloquium},
such as twisted moiré materials which host nearly flat bands with general quantum geometry.

Moreover, the localization criticality of Chern insulator has theoretical significance in its own right.
For disordered Chern insulators with Chern number $\abs{C}=1$ and no additional symmetries,
their localization transitions
are classified into the unitary class~\cite{huckestein1995scaling,evers2008anderson}, the same class as IQHT,
which breaks time-reversal invariance and includes a topological $\theta$ term in its critical theory.
It has long been conjectured~\cite{ludwig1994integer} that
the fixed point of disordered Dirac fermions in 2D also governs the criticality of IQHT.
The tunable quantum geometry in ideally flat Chern band
then provides a brand new degree of freedom for examining the localization transition in the unitary class.
In our numerical studies, a universal critical exponent $\nu=2.15(1)\sim2.19(2)$ is found
when the quantum metric is Dirac-like, i.e.,
when $g_{\mu\nu}(k)$ is enhanced and asymptotically singular at the Dirac points.
Conversely, when the quantum metric develops a singular line,
our data are consistent with a crossover toward the orthogonal class $\nu=\infty$,
giving rise to a non-universal and increasing effective critical exponent although the band gap remains open.
This crossover from universal to non-universal regime is dominated entirely by the quantum metric,
which may shed new light on the long-standing discrepancies in reported IQHT critical exponents
across experiments and numerical studies, as well as across individual numerical reports.

{\color{myblue}\it Transfer matrix method in the maximally localized hybrid Wannier basis.---}
To realize an ideally flat Chern band,
we adopt the celebrated $\pi$-flux model~\cite{neupert2011fractional}
on the square lattice as the parent Hamiltonian,
which is one of the earliest predicted fractional Chern insulators at zero magnetic field.
The local Hilbert space is spanned by two sublattice orbitals for spinless fermions,
and the two-band Bloch Hamiltonian is defined as
$
    H(k)=\sum_{i=x,y,z} h_{i,k} \tau_i
$,
with
$
    h_{x,k} + i h_{y,k} = -t_1 e^{i\phi} [1+e^{i(k_x+k_y)}] - t_1 e^{-i\phi} [e^{ik_x}+e^{ik_y}]
$
and
$
    h_{z,k} = -2 t_2 (\cos k_x - \cos k_y)
$.
$\tau_i$ are the Pauli matrices
and $t_1,t_2\geqslant0$ denote the inter- and intra-sublattice hoppings respectively,
as illustrated in Fig.~S1 of Ref.~\cite{supplement}.
We fix $\phi=\pi/4$ to yield the staggered $\pm\pi$ flux through plaquettes.
A nonzero $\phi$ breaks the time-reversal symmetry and drives the model into a topological phase with Chern number $\abs{C}=1$ when the band gap is open.
The topological transition to a gapless state occurs at $t_2/t_1=0$ and $\infty$.
In addition, the absence of $\tau_0$ term in the Hamiltonian ensures a particle-hole symmetric band dispersion,
and both bands become optimally flat when $\sqrt{2}t_2/t_1=1$,
as shown in Fig.~\ref{fig:fig1}(b).
An exactly flat Chern band for general $t_2/t_1$
is constructed by manually flattening $H(k)$ while preserving its Bloch eigenstates and hence its quantum geometry, i.e., through
$
    H_\text{flat}(k) = H(k)/\abs{\varepsilon_k} = \sum_{\alpha=\pm1}\alpha\ket[false]{u_{\alpha,k}}\bra[false]{u_{\alpha,k}}
$.
It is proved in Ref.~\cite{neupert2011fractional} that the flattened model preserves locality
in the sense that the effective hopping amplitudes decay exponentially with distance.

We note that the locality of the chosen basis is essential for transfer matrix calculations,
as it allows $H_\text{flat}$ to be represented in a spatial block structure up to a controlled spatial truncation.
Most naturally, $H_\text{flat}$ under the atomic orbital basis is localized in both directions,
while its localization property is not optimized, leading to substantial numerical inefficiency in practice.
More importantly, the decay of hopping amplitudes in atomic orbital basis
is not related to any intrinsic quantum geometric length in an obvious way.
To overcome these drawbacks, we switch to the hybrid Wannier basis~\cite{marzari2012maximally,supplement},
\begin{equation}\label{eq:hybrid-wannier-basis}
    \ket{W_{k_y,x}} = \frac{1}{\sqrt{L}} \sum_{k_x} e^{-ik_xx} e^{i\phi_{k_x,k_y}} \ket{\psi_{k_x,k_y}},
\end{equation}
which is localized in $x$ direction and extended in $y$,
mimicking the lowest Landau level wavefunction in the Landau gauge.
Throughout this work, we focus on the isolated lower band of the $\pi$-flux model with Chern number $C=1$.
We neglect band mixing and hence the band indices are omitted in Eq.~\eqref{eq:hybrid-wannier-basis} and hereafter.
A square lattice with length $L$ in $x$ and width $M$ in $y$ is adopted under the periodic boundary condition.
The Bloch wavefunction $\ket[false]{\psi_{k_x,k_y}}$ is related to the cell-periodic part $\ket[false]{u_{k_x,k_y}}$
through the Bloch theorem
$
    \ket[false]{\psi_k}_{\alpha,r} = e^{ikr} \ket[false]{u_k}_\alpha
$.
The phase $\phi_{k_x,k_y}$ in Eq.~\eqref{eq:hybrid-wannier-basis}
highlights the gauge freedom of Wannier basis,
which dramatically affects the localization property of Wannier states~\cite{marzari1997maximally}.
It is well established that in (quasi) one dimension (1D), the maximally localized (hybrid) Wannier function
is determined by adopting the parallel transport gauge~\cite{marzari1997maximally,qi2011generic,lee2013pseudopotential},
\begin{subequations}\begin{align}
    \phi_{k_x,k_y} &= \int_0^{k_x} \mathrm{d}k_x A_x - k_x \phi_{B}(k_y),\\
    \phi_{B}(k_y) &= \frac{1}{2\pi}\int_0^{2\pi}\mathrm{d}{k_x}A_{x},
\end{align}\end{subequations}
where $A_x = i\expt[false]{u_k}{\partial_{k_x}u_k}$
is the $k_x$-component of the Berry connection in 2D,
and $\phi_B(k_y)$ is the Berry phase accumulated along $k_x$ direction, divided by $2\pi$.
In this gauge, the hybrid Wannier functions are exponentially localized along $x$ and the Wannier centers are given by
$
    r_{k_y,x} = x + \phi_{B}(k_y)
$
for $x$ in the bulk.
$\phi_B(k_y)$ thus also denotes the charge polarization away from the lattice site $x$.
The spatial spread of Wannier basis in quasi 1D is quantified by the Wannier variance
$
    \Omega(k_y) = \expt[false]{W_{k_y,x}}{x^2}{W_{k_y,x}} - \expt[false]{W_{k_y,x}}{x}{W_{k_y,x}}^2
$.
It has been proved in the pioneering work~\cite{marzari1997maximally} that under the parallel transport gauge,
the Wannier variance $\Omega(k_y)$ is reduced to a minimal and gauge-invariant value $\Omega_I(k_y)$
which is a quantum geometric quantity,
\begin{equation}\label{eq:omega-min}
    \min\left[\Omega(k_y)\right] = \Omega_I(k_y) = a \int \frac{\mathrm{d}k_x}{2\pi}\ g_{xx}(k),
\end{equation}
with $a$ the lattice constant and $g_{\mu\nu}$ the quantum metric.
Therefore, we define the associated quantum geometric length
\begin{equation}\label{eq:geometric-length}
    \xi_\text{geo.}(k_y) \overset{!}{=} \sqrt{\Omega_I(k_y)},
\end{equation}
which \textit{exactly} characterizes the spatial spread of maximally localized hybrid Wannier basis with momentum $k_y$.
We demonstrate the exponential decay of maximally localized hybrid Wannier basis in Fig.~\ref{fig:fig1}(d)
and the quantum geometric length $\xi_\text{geo.}(k_y)$ in Fig.~\ref{fig:fig1}(e).

Employing the maximally localized hybrid Wannier basis
makes the role of quantum geometric length explicit and, as shown below,
enables an efficient transfer matrix calculation of localization length.
In this work, on-site and uncorrelated disorder is considered,
$
    V=\sum_{r,\alpha}V_{r,\alpha}c^\dag_{r,\alpha}c_{r,\alpha}
$,
where $V_{r,\alpha}$ are independent real random variables.
We incorporate either the Anderson disorder, where $V_{r,\alpha}$ are drawn from
a uniform distribution $\text{Uniform}(-W,W)$ with $W$ the disorder strength,
or the white-noise disorder generated from a normal distribution $\text{Normal}(0,W/\sqrt{3})$.
These two setups are constructed to yield disorder distributions
with zero mean and the same standard deviation $W/\sqrt{3}$.
Since we focus on the isolated lower band, the band mixing is strictly prohibited;
this assumption is justified in the weak-disorder regime
where the disorder strength $W$ is significantly smaller than the band gap.
As a result, $W$ serves as the only energy scale of the system and is set to unity once for all.
We then formulate the disordered flat-band Hamiltonian
in the lower-band hybrid Wannier subspace as
\begin{equation}
    H_\text{flat} = \sum_{xx' k_yk_{y'}} \left[V_{x,x'}\right]_{y,y'} \ket[false]{W_{k_y,x}} \bra[false]{W_{k_{y'},x'}},
\end{equation}
with the flat-band energy set to zero.
$[V_{x,x'}]_{y,y'}=\expt[false]{W_{k_y,x}}{V}{W_{k_{y'},x'}}$
denotes the matrix elements under the maximally localized hybrid Wannier basis.
Because hybrid Wannier states are exponentially localized along $x$,
elements of $V_{x,x'}$ also decay exponentially with $\abs{x-x'}$.
Therefore it is safe to truncate $V_{x,x'}$ for large separation $\vert x-x'\vert>l_0$
such that the Hamiltonian can be divided into a block form suitable for transfer matrix calculations.
In practice, we have chosen $l_0$ such that $\Vert V_{x,x+l_0}\Vert/\Vert V_{x,x}\Vert \lesssim 10^{-3}$,
and tested the convergence of our results with respect to varied $l_0$~\cite{supplement}.
For transfer matrix calculations,
we consider a stripe geometry ($L\gg M$) with fixed aspect ratio $L/M=64$.
The system width $M$ is varied up to 64 and extended to $M=96$ for specific value of $t_2/t_1$.
For the transfer matrix algorithm under hybrid Wannier basis and necessary implementation details,
readers may refer to Ref.~\cite{supplement}.

\begin{figure}[htbp]
    \centering\hspace{0cm}
    \includegraphics[width=1\columnwidth]{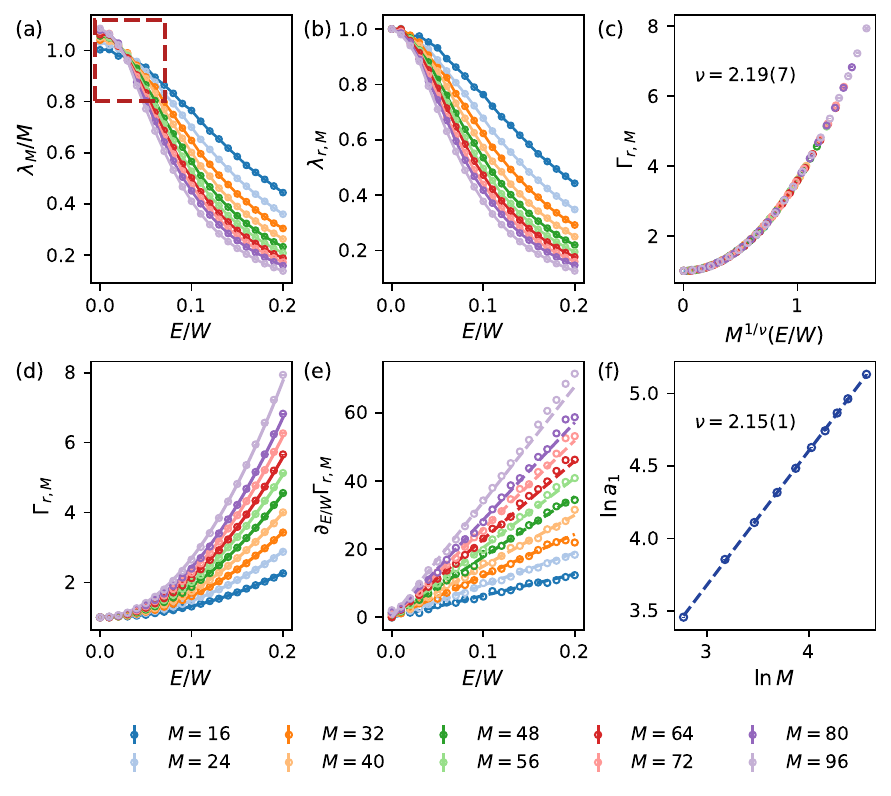}
    \caption{%
        (a) Localization length $\lambda_M$ with white-noise disorders and $\sqrt{2}t_2/t_1=0.5$.
        (b) Reduced localization length $\lambda_{r,M}(x)=\Gamma_{r,M}(x)^{-1}$ under the factorization ansatz.
        (c) Data collapse of $\Gamma_{r,M}$ according to the single-parameter scaling Eq.~\eqref{eq:single-parameter-scaling}.
        (d)-(f) Minimal fitting scheme (MFS) of extracting the critical exponent,
        where we fit $\Gamma_{r,M}$ as $1+a_1(E/W)^2$ such that $\ln a_1\sim \frac{2}{\nu} \ln M$.
    }
    \label{fig:fig2}
\end{figure}

{\color{myblue}\it Critical exponent of the unitary class.---}
Before we can extract the localization length in the thermodynamic limit,
the critical exponent of localization transition should first be determined.
In transfer matrix calculations, the localization lengths $\lambda_M(E)$ are measured
for systems with finite width $M$.
To estimate the critical exponent as $M\to\infty$,
we follow the standard two-parameter scaling analysis,
which is well established for IQHT and includes an, possibly marginally, irrelevant scaling field~\cite{slevin2009critical,nuding2015localization},
\begin{equation}\label{eq:scaling}
    \Gamma_M(x) \overset{!}{=} \frac{M}{\lambda_M(x)} = \Gamma\left(M^{1/\nu}x, f(M)\right),
\end{equation}
where $\Gamma$ is the collapse function and $\nu>0$ the critical exponent to be determined.
Eq.~\eqref{eq:scaling} is valid in the regime $x\ll1$
where the relevant and (marginally) irrelevant scaling fields are expanded
to the first and zeroth order in $x=E/W$ respectively.
For an irrelevant field, $f(M)=M^{-y}$ with $y>0$;
for a marginally irrelevant field, the logarithmic correction is captured by $f(M)=(\ln M)^{-p}$ with $p>0$.
In our case, incorporating the irrelevant or marginally irrelevant contribution is essential.
This necessity is evidenced in Fig.~\ref{fig:fig2}(a),
where $\lambda_M/M$ for varied $M$ fail to converge as $x\to0$ and a kink appears at finite $x$.
Also, because the disorder distributions involve balanced attractive and repulsive scatterers,
the localization lengths are symmetric about the band center $E_c=0$; we therefore scan only positive energies.

\begin{figure}[htbp]
    \centering\hspace{-.75cm}
    \includegraphics[width=.95\columnwidth]{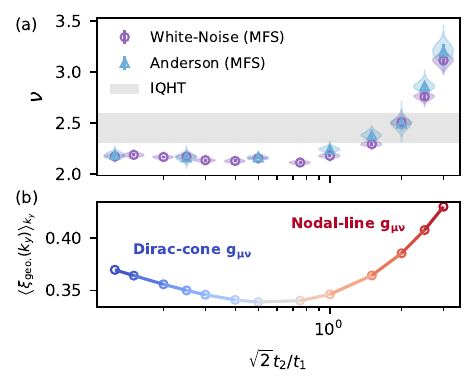}
    \caption{%
        (a) Critical exponent $\nu$ for varied $t_2/t_1$, extracted through MFS.
        The distributions of $\nu$, i.e. the shadowed violins, are estimated from bootstrap resampling.
        The grey ribbon indicates the reported range of IQHT critical exponent,
        $\nu_\text{IQHT}=2.3\sim2.6$, as discussed in the main text.
        (b) Evolution of quantum geometric length for varied $t_2/t_1$.
    }
    \label{fig:fig3}
\end{figure}

\begin{figure*}[tp]
    \centering\hspace{-.5cm}
    \includegraphics[width=.9\linewidth]{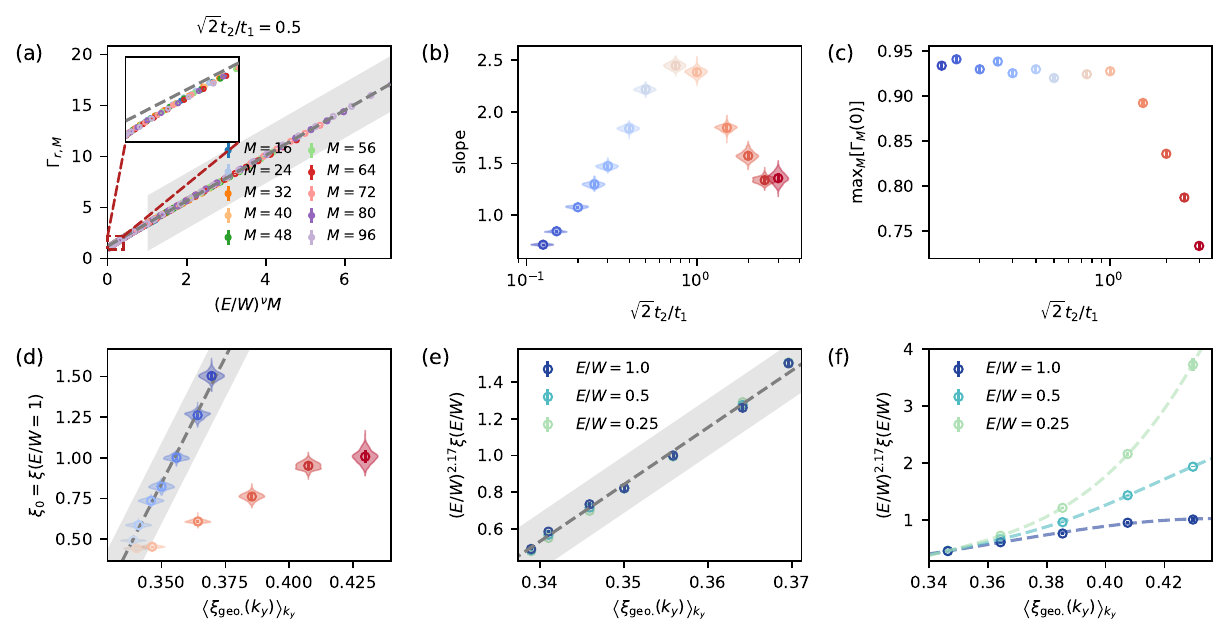}
    \caption{%
        Localization length $\xi=\xi_0 (E/W)^{-\nu}$.
        (a) Collapsed $\Gamma_{r,M}$ as a function of $M(E/W)^\nu$.
        The grey ribbon marks the range of data used for the linear fitting to estimate $\xi_0$.
        (b) Fitted slope of the collapse function $\Gamma_r$.
        (c) $\Gamma_M(0)$ for the largest accessible system size $M$ at each $t_2/t_1$.
        (d)-(f) Localization length $\xi(E/W)$ versus momentum-averaged quantum geometric length $\langle\xi_\text{geo.}(k_y)\rangle_{k_y}$.
        Localization lengths at varied energies, scaled by an energy-dependent factor $(E/W)^{2.17}$,
        are shown for the (e) universal and (f) non-universal regime of critical exponent.
    }
    \label{fig:fig4}
\end{figure*}

In general, one has to expand the right-hand side of Eq.~\eqref{eq:scaling} into a polynomial,
whereby $\nu$ is determined through fitting numerous expansion parameters.
In practice, we find our data well fitted into a factorization ansatz,
\begin{equation}
    \Gamma_M(x) = \Gamma_0\left(M^{1/\nu}x\right)\ \Gamma_1\left(f(M)\right),
\end{equation}
where the relevant and irrelevant scaling fields are factorized.
We note that a similar ansatz was also used to analyze the scaling of longitudinal conductance of IQHT~\cite{dresselhaus2022scaling}.
Under the factorization ansatz, if we define
$
    \Gamma_{r,M}(x) = \Gamma_M(x)/\Gamma_M(0)
$,
which eliminates the irrelevant scaling $\Gamma_1$,
then $\Gamma_{r,M}(x)$ should follow a single-parameter scaling behavior
\begin{equation}\label{eq:single-parameter-scaling}
    \Gamma_{r,M}(x) = \Gamma_{r}\left(M^{1/\nu}x\right),
\end{equation}
as illustrated in Fig.~\ref{fig:fig2}(b)(c).
Since $\Gamma_1$ has been completely factored out,
it is not our focus here to determine whether the irrelevant scaling field is marginal or not.
The critical exponent $\nu$ is then estimated either from the data collapse shown in Fig.~\ref{fig:fig2}(c)
or a minimal fitting scheme (MFS) as in Fig.~\ref{fig:fig2}(d)-(f).
In the latter, $\Gamma_{r,M}(x)$ is fitted for small $x$ as $1+a_1 x^2$,
and $\nu$ can be extracted from the relation $\ln a_1\sim \frac{2}{\nu} \ln M$.

We show in Fig.~\ref{fig:fig3} the critical exponent $\nu$ extracted from MFS,
together with the momentum-averaged quantum geometric length defined in Eq.~\eqref{eq:geometric-length},
as a function of $t_2/t_1$.
Employing data collapse generally yields consistent critical exponents,
though subject to larger statistical uncertainties.
Extensive data regarding the finite-size scaling, critical exponent values, and the benchmark between data collapse and MFS results
are provided in Ref.~\cite{supplement}.
In Fig.~\ref{fig:fig3}(a), a universal critical exponent of $\nu=2.15(1)\sim2.19(2)$ is revealed at small $t_2/t_1$.
Notably, $\nu$ undergoes a prominent flow toward higher values when $\sqrt{2}t_2/t_1\gtrsim1$.
This trend remains consistent across different types of disorder, suggesting a degree of generality against microscopic details.
Moreover, it is unlikely to be attributed solely to finite-size effects
since the quantum geometric length, or the characteristic length scale of the system, diverges at both $t_2/t_1=0$ and $+\infty$,
yet the non-universal behavior of critical exponent emerges exclusively on the large $t_2/t_1$ side.
Instead, we find that the crossover to the non-universal regime is closely correlated with the evolution of quantum metric.
As shown in Fig.~\ref{fig:fig3}(b) and Fig.~S2 in Ref.~\cite{supplement},
the $\pi$-flux Hamiltonian at small $t_2/t_1$ describes massive Dirac fermions,
with quantum metric weight concentrated near Dirac points.
However, at large $t_2/t_1$, the diagonal $\tau_z$ term dominates;
in the $t_1=0$ limit this yields a gapless line,
and correspondingly the quantum metric is prominently distributed along nodal lines in momentum space.

The observation that the universal regime of critical exponent features a Dirac-like quantum metric
suggests a direct relevance to Ludwig's seminal conjecture~\cite{ludwig1994integer},
stating that the IQHT and disordered Dirac fermions in 2D are governed by the same fixed point.
Although the Chern band here is exactly flat, it inherits the quantum metric of Dirac fermions,
and hence we expect it hosts a stable fixed point of disordered Dirac fermion within the unitary class.
This Dirac fixed point with universal critical exponent is also supported by
a stable $\Gamma_M(0)$ in this regime as shown in Fig.~\ref{fig:fig4}(c).
Furthermore, we propose that the non-universal regime of critical exponent arises from
\textit{the crossover to the orthogonal class fixed point} at $t_2/t_1=+\infty$.
At $t_2/t_1=+\infty$ or $t_1=0$, the localization transition falls into the orthogonal class
as the gap is closed and the time-reversal symmetry is recovered.
The critical exponent of orthogonal class is evaluated in dimension $d=2+\epsilon$~\cite{wegner1980disordered,hikami1981anderson,wegner1989fourlooporder,evers2008anderson}
as $\nu=\epsilon^{-1}+O(\epsilon^2)$,
and hence in exact 2D the critical exponent formally diverges.
At large $t_2/t_1$ with a nodal-line quantum metric,
our numerics are qualitatively consistent with an effective critical exponent
driven to larger values by the orthogonal class fixed point,
even though the topological band gap remains open at finite $t_2/t_1$.

In Fig.~\ref{fig:fig3}(a), we also marked the range of reported IQHT critical exponents in the literature.
Latest IQHT experiments reported a critical exponent of 2.4
in both 2DEG~\cite{wei1988experiments,li2005scaling,li2009scaling}
and graphene devices~\cite{kaur2024universality}.
On the numerical side, IQHT criticality has been extensively examined
by considering either the Chalker-Coddington (CC) network model~\cite{chalker1988percolation,kramer2005random,
slevin2009critical,amado2011numerical,obuse2012finitesize,nuding2015localization,dresselhaus2022scaling}
or electrons under the magnetic field in the continuum~\cite{huckestein1990oneparameter,
liu1994universal,huo1992current,zhu2019localizationlength},
on the lattice~\cite{puschmann2019integer,zhu2019localizationlength},
and also in a dual composite-fermion representation~\cite{huang2021numerical}.
The main numerical approaches include
transfer matrix calculations~\cite{slevin2009critical,amado2011numerical,
obuse2012finitesize,nuding2015localization,huckestein1990oneparameter,liu1994universal,puschmann2019integer},
the scaling of current-carrying states~\cite{huo1992current,zhu2019localizationlength},
and the scaling of longitudinal conductance~\cite{dresselhaus2022scaling}.
As reviewed in Ref.~\cite{dresselhaus2021numerical},
the reported critical exponents $\nu$ among these studies
range from 2.3 to 2.6 with errorbars of order $10^{-2}$,
therefore indicating a remarkable discrepancy across models and analyses.
In comparison, the universal critical exponent of Dirac fixed point observed in our work
lies below the commonly quoted IQHT window.
Our findings on the quantum-metric-induced crossover from a universal to non-universal regime
hence suggest a novel possibility:
since LLs host ideally flat quantum metric distinct from a Dirac-like one,
IQHT may reside within the crossover regime and thus manifest a non-universal critical behavior.
Correspondingly, deviations from the ideal LL limit may drive a drift of the effective critical exponent.

On the other hand, the localization of Chern bands has also been studied in Refs.~\cite{onoda2003quantized,chang2016observation,
mildner2023topological,ivaki2020criticality,bera2024quantum},
and floating critical exponents were suggested in
the Haldane model~\cite{mildner2023topological},
an amorphous Chern model~\cite{ivaki2020criticality},
and Dirac fermions~\cite{sbierski2021criticality}.
In Ref.~\cite{sbierski2021criticality},
the disorder transition in Dirac fermions was examined by tuning the Dirac mass at certain fixed energy,
and the critical exponent was found energy-dependent, evolving from 2.33(3) at $E=0$ to 2.53(2) at $E=0.7$.
We note that the critical regime in these studies~\cite{mildner2023topological,ivaki2020criticality,sbierski2021criticality}
is restricted by the significant bandwidth.
Consequently, the role of quantum geometric length can be obscured due to the presence of additional length scales.
Also, there has been recent experimental progress~\cite{kawamura2020current}
in measuring the critical scaling of a quantum anomalous Hall insulator,
yet the reported critical exponent $\nu=2.8(3)$ is subject to considerable statistical uncertainty.

{\color{myblue}\it Quantum geometric localization length.---}
To proceed, we determine the localization length $\xi=\xi_0\abs{x}^{-\nu}$, especially $\xi_0$,
in the thermodynamic limit and confirm its quantum geometric origin.
Note that the single-parameter scaling in Eq.~\eqref{eq:single-parameter-scaling}
describes a scaling behavior with the localization length itself as the scaling variable.
For sufficiently large $M$, $\lambda_M(x)$ converges to $\xi(x)$
and hence Eq.~\eqref{eq:single-parameter-scaling} yields
$
    M/\left[\Gamma_M(0)\xi(x)\right] = \Gamma_r\left(M\xi_0/\xi(x)\right)
$.
This implies the asymptotic form of collapse function $\Gamma_r(z)$ at $z\gg1$,
\begin{equation}
    \Gamma_r(z) = \frac{z}{\xi_0\lim_{M\to\infty}\Gamma_M(0)} + O(1).
\end{equation}
As a result, $\xi_0$ can be extracted by first collapsing $\Gamma_{r,M}(x)$ against $Mx^\nu$
and then performing a linear fit to the collapsed curve at large $Mx^\nu$.
This procedure is shown in Fig.~\ref{fig:fig4}(a), where a prominent linear relation is observed.
In practice, extrapolating $\Gamma_M(0)$ to the thermodynamic limit is numerically demanding.
Therefore we estimate $\xi_0$ using $\Gamma_M(0)$ at the largest accessible $M$.

We show the fitted slope of $\Gamma_r$ in Fig.~\ref{fig:fig4}(b) and $\Gamma_M(0)$ in Fig.~\ref{fig:fig4}(c),
both of which exhibit qualitatively distinct behavior in the universal (blue) and crossover (red) regimes of critical exponent.
The localization lengths $\xi_0$ are plotted in Fig.~\ref{fig:fig4}(d)
against the momentum-averaged quantum geometric length.
Two distinct branches are observed,
corresponding to the universal regime associated with Dirac fixed point and the crossover regime respectively.
The distributions of the fitted slope and $\xi_0$ are obtained via bootstrap resampling,
from which their standard deviations are estimated.
Within the universal regime, localization length $\xi_0$ exhibits a clear linear dependence on the quantum geometric length.
This scaling is also evidenced by $\xi(E/W)$ at varied energies, as shown in Fig.~\ref{fig:fig4}(e).
By contrast, because the critical exponent is generally non-universal in the crossover regime,
$\xi(E/W)$ in Fig.~\ref{fig:fig4}(f) manifest energy-dependent behavior,
while still maintaining a positive correlation with the quantum geometric length.

Furthermore, the quantum geometric localization length predicts a quantum geometric mobility edge.
For a finite system, delocalized states exist at energies where
the localization length $\xi(E)$ exceeds the system size $M$.
The associated critical energy, the mobility edge, $E_m$ is given by
\begin{equation}
    E_m/W = \left(\xi_0/M\right)^{1/\nu}.
\end{equation}
Therefore, an ideally flat Chern band with larger quantum geometric length experiences weaker localization,
and correspondingly exhibits both a larger $\xi_0$ and a larger mobility edge.
The quantum geometric effects on the mobility edge shall be reflected in
the sharpness of plateau transition and the broadening of longitudinal resistivity $\rho_{xx}$
in flat quantum anomalous Hall systems.
Recent moiré materials with tunable quantum geometry~\cite{adak2024tunable} may therefore
serve as a promising platform to confirm our findings.
We also note that, based on the DFT calculations in Refs.~\cite{wang2024fractional,xu2024maximally,wang2023topology},
the characteristic quantum geometric length in twisted bilayer $\text{Mo}\text{Te}_2$ at twist angle $\theta=3.89^\circ$
is among 2.1~nm to 2.5~nm.
This scale is comparable to the moiré period and much larger than the bare lattice constant $a_0\approx0.35\text{ nm}$.


\newcommand{\nocontentsline}[3]{}
\let\origcontentsline\addcontentsline
\newcommand\stoptoc{\let\addcontentsline\nocontentsline}
\newcommand\resumetoc{\let\addcontentsline\origcontentsline}

\stoptoc

\begin{acknowledgments}
{\color{myblue}\it Acknowledgments.---}
We thank Xin Wan and Wei Zhu for inspiring discussions.
This work is supported by
the National Key R\&D Program of China (Grant No.~2022YFA1403402),
the National Natural Science Foundation of China (Grant No.~12174068),
the Science and Technology Commission of Shanghai Municipality (Grant Nos.~24LZ1400100 and 23JC1400600),
and the Shuguang Program of Shanghai Education Development Foundation and Shanghai Municipal Education Commission.
The authors also acknowledge \href{https://www.paratera.com}{Beijing PARATERA Tech Co., Ltd.}
and the CFFF platform of Fudan University
for providing the computational resources used in this work.
\end{acknowledgments}

\bibliographystyle{apsrev4-2}
\bibliography{ref.bib}

\resumetoc

\newpage
\clearpage
\onecolumngrid

\begin{center}
\textbf{
Supplementary Material for\\
``Quantum geometric localization length and localization criticality in an ideally flat Chern band"
}
\end{center}

\setcounter{equation}{0}
\setcounter{figure}{0}
\setcounter{table}{0}
\setcounter{page}{1}

\makeatletter

\renewcommand{\thetable}{S\arabic{table}}
\renewcommand{\theequation}{S\arabic{equation}}
\renewcommand{\thefigure}{S\arabic{figure}}
\renewcommand{\bibnumfmt}[1]{[S#1]}
\renewcommand{\citenumfont}[1]{S#1}
\setcounter{secnumdepth}{3}

\tableofcontents

\section{$\pi$-flux model and quantum geometry}
\begin{figure}[htbp]
    \centering\hspace{0cm}
    \includegraphics[width=.3\columnwidth]{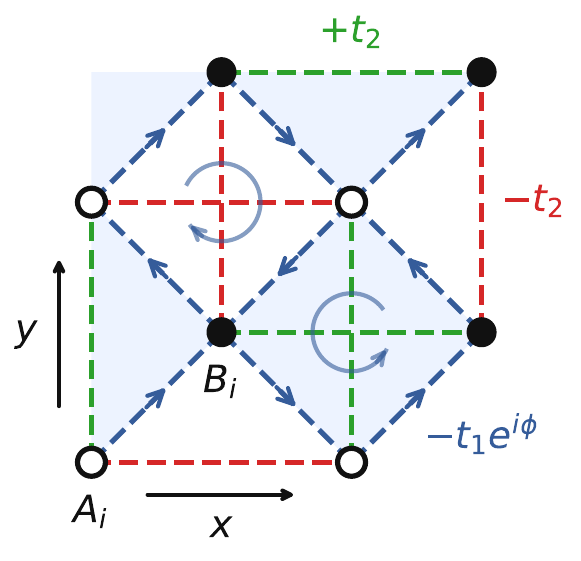}
    \caption{%
        $\pi$-flux model on the square lattice.
    }
    \label{fig:figs1}
\end{figure}

We transcribe here the Bloch Hamiltonian of the $\pi$-flux model~\cite{neupert2011fractional} on the square lattice as
\begin{equation}
    \mathcal{H}(k) = \sum_{i=x,y,z} h_{i,k} \tau_i = 
    \begin{pmatrix}
        h_{z,k} & h_{x,k} - i h_{y,k}\\
        h_{x,k} + i h_{y,k} & -h_{z,k}\\
    \end{pmatrix},
\end{equation}
with
\begin{subequations}\label{eq:pi-flux-hamil}
\begin{align}
    h_{z,k} &= -2 t_2 (\cos k_x - \cos k_y),\\[5pt]
    h_{x,k} + i h_{y,k} &= -t_1 e^{i\phi} \left(1+e^{i(k_x+k_y)}\right) - t_1 e^{-i\phi} \left(e^{ik_x}+e^{ik_y}\right).
\end{align}
\end{subequations}
As illustrated in Fig.~\ref{fig:figs1}, each unit cell involves two sublattice sites $A_i$ and $B_i$.
The nearest-neighbor hopping $t_1$ couples sites of $A$ and $B$ sublattices,
and accumulates a phase $\phi=\pi/4$ along the direction indicated by the blue arrows,
thereby contributing to the staggered $\pm\pi$ flux.
The next-nearest-neighbor hopping $t_2$ connects sites of the same sublattice
and further carries opposite signs between $x$/$y$ direction and $A$/$B$ sublattice.

Define the \textit{Hamiltonian Bloch vector} $\bm{h}_k=(h_{x,k},h_{y,k},h_{z,k})$.
The eigenenergies and eigenstates are expressed as
\begin{subequations}
\begin{align}
    \varepsilon_{\pm,k} &= \pm \lvert\bm{h}_k\rvert,\\[8pt]
    \ket{u_{+,k}} &=
    \begin{pmatrix}
    e^{-i\phi_k/2} \cos\frac{\theta_k}{2}\\[4pt]
    e^{+i\phi_k/2} \sin\frac{\theta_k}{2}
    \end{pmatrix},
    \quad
    \ket{u_{-,k}} =
    \begin{pmatrix}
    e^{-i\phi_k/2} \sin\frac{\theta_k}{2}\\[4pt]
    -e^{+i\phi_k/2} \cos\frac{\theta_k}{2}
    \end{pmatrix},
\end{align}
\end{subequations}
where $\phi_k=\arg(h_{x,k}+ih_{y,k})$ and $\cos\theta_k=h_{z,k}/\abs{\bm{h}_k}$ for nonzero $\abs{\bm{h}_k}$.

\subsection{Quantum geometry basics}
The gauge-invariant \textit{quantum geometric tensor}~\cite{cheng2013quantum,graf2021berry}
for an isolated band is defined as
\begin{equation}
    Q_{\mu\nu}(\lambda) \stackrel{!}{=} \expt{\partial_\mu\psi(\lambda)}{\partial_\nu\psi(\lambda)} - \expt{\partial_\mu\psi(\lambda)}{\psi(\lambda)} \expt{\psi(\lambda)}{\partial_\nu\psi(\lambda)},
\end{equation}
where $\partial_\mu=\partial/\partial_{\lambda_\mu}$.
The real part
$
    g_{\mu\nu} = \text{Re}\ Q_{\mu\nu}
$
is known as the \textit{Fubini-Study (FS) metric},
and the imaginary part
$
    \sigma_{\mu\nu} = \text{Im}\ Q_{\mu\nu}
$
is related to the \textit{Berry curvature}.
By construction, $g_{\mu\nu}$ is a symmetric tensor and $\sigma_{\mu\nu}$ an antisymmetric one.
The FS metric $g_{\mu\nu}$ serves as the metric tensor for measuring the invariant distance of two quantum states
in the parameter space of $\lambda_\mu$.
Recall the definition of Berry connection and Berry curvature,
\begin{subequations}\begin{align}
    A_\mu &= i\expt{\psi(\lambda)}{\partial_\mu\psi(\lambda)},\\[5pt]
    F_{\mu\nu} &= \partial_\mu A_\nu - \partial_\nu A_\mu = i \left[
        \expt{\partial_\mu\psi(\lambda)}{\partial_\nu\psi(\lambda)}- \expt{\partial_\nu\psi(\lambda)}{\partial_\mu\psi(\lambda)}
    \right]
    = i \left[Q_{\mu\nu} - Q_{\nu\mu}\right],\label{eq:berry-curvature}
\end{align}\end{subequations}
where in the last step of Eq.~\eqref{eq:berry-curvature}
we have noted $\text{Re}\expt[false]{\psi}{\partial_\mu\psi} = 0$.
It is immediately realized that $F_{\mu\nu} = -2\sigma_{\mu\nu}$,
and the quantum geometric tensor can be expressed in terms of the FS metric and Berry curvature as
\begin{equation}
    Q_{\mu\nu} = g_{\mu\nu} - \frac{i}{2} F_{\mu\nu}.
\end{equation}

\begin{figure}[htbp]
    \centering
    \includegraphics[width=1\columnwidth]{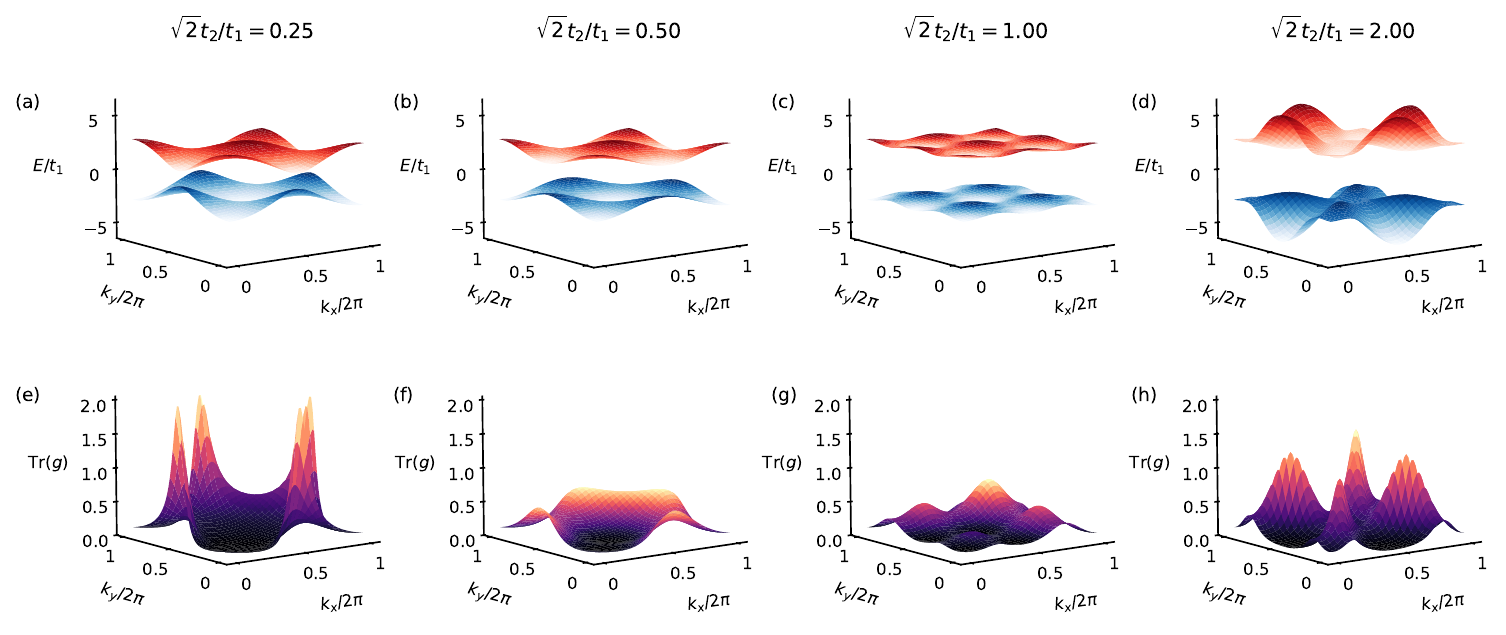}
    \caption{%
        Evolution of (a)-(d) band dispersion and (e)-(h) lower-band quantum metric of the $\pi$-flux model.
    }
    \label{fig:figs2}
\end{figure}

Based on these definitions, for a general two-band model,
$g_{\mu\nu}$ and $F_{\mu\nu}$ are conveniently expressed with $\bm{h}_k$ as~\cite{graf2021berry}
\begin{subequations}
\begin{align}
    g^\pm_{\mu\nu}(k) &= \frac{1}{4\lvert\bm{h}\rvert^2} \left[\bm{h}^\mu\cdot\bm{h}^\nu - \frac{\left(\bm{h}\cdot\bm{h}^\mu\right)\left(\bm{h}\cdot\bm{h}^\nu\right)}{\lvert\bm{h}\rvert^2}\right],\\[10pt]
    F^\pm_{\mu\nu}(k) &= \mp\frac{1}{2\lvert\bm{h}\rvert^3} \bm{h}\cdot\left(\bm{h}^\mu\times\bm{h}^\nu\right),
\end{align}
\end{subequations}
where $\bm{h}^\mu$ is the shorthand for $\partial_\mu\bm{h}$
and we specify the parameter space as the first Brillouin zone.
In Fig.~\ref{fig:figs2}, we plot the band dispersion and lower-band quantum metric of the $\pi$-flux model.
The integral of Berry curvature over the Brillouin zone yields the first Chern number, which is a topological invariant,
\begin{equation}
    C_\pm = \frac{1}{2\pi} \int \mathrm{d}^2k\ F^\pm_{xy}(k)\ \in\ \mathbb{Z}.
\end{equation}
For the lower band of $\pi$-flux model, its Chern number $C_{-}=1$.

Next, we introduce some rigorous inequalities~\cite{roy2014band,ozawa2021relations} that
impose constraints on the quantum geometry in 2D systems,
\begin{equation}\label{eq:inequalities}
    \text{Tr}\ g^\alpha(k) \geqslant 2\sqrt{\text{det}\ g^\alpha(k)} \geqslant |F^\alpha_{xy}(k)|,
\end{equation}
where $\alpha$ denotes the band index.
For a two-band model defined in a two-torus parameter space, as in our case,
Ref.~\cite{ozawa2021relations} proves that
the second inequality is always saturated, yielding an exact relation
$\sqrt{\text{det}\ g^\alpha(k)} = \frac{1}{2}|F^\alpha_{xy}(k)|$.
In addition, the first inequality is saturated if and only if
the local $g^\alpha(k)$ is proportional to identity,
\begin{equation}
    \text{Tr}\ g^\alpha(k) = 2\sqrt{\text{det}\ g^\alpha(k)} \quad \leftrightarrow \quad g^\alpha(k) \propto \mathds{1}_{2\times2},
\end{equation}
which is a direct consequence of $g^\alpha(k)$ being a positive semi-definite matrix.
These inequalities will be useful when later we compare different definitions of quantum geometric lengths.
As a special example, the quantum geometry of $r$-filled Landau levels~\cite{ozawa2021relations} is
\begin{equation}
    g = \frac{r}{2\abs{B}}
    \begin{pmatrix}
    1 & 0\\
    0 & 1
    \end{pmatrix}, \quad
    F_{xy} = -\frac{r}{B}.
\end{equation}
Both of them are completely flat in the Brillouin zone, and all of the above inequalities are saturated.

\subsection{Quantum geometric lengths}
\label{sec:geometric-lengths}
In this work, we propose that the localization length in a flat Chern band has a quantum geometric origin,
and is governed by a quantum geometric length related to the quantum metric.
It will be shown below that such a quantum geometric length emerges as
the characteristic spatial spread of the maximally localized hybrid Wannier basis.

For an isolated band, the spatial spread of Wannier basis is measured by the Wannier variance
$
    \Omega = \expt[false]{W_R}{r^2}{W_R} - \expt[false]{W_R}{r}{W_R}^2
$,
which can be separated into an intrinsic gauge-invariant part $\Omega_I$ and a gauge-dependent part $\widetilde{\Omega}$.
In general $d$ dimension,
the intrinsic Wannier variance $\Omega_I$ is a quantum geometric quantity~\cite{marzari1997maximally},
\begin{equation}
    \Omega_I = V_\text{cell} \int \frac{\mathrm{d}^dk}{(2\pi)^d}\ \text{Tr}\ g(k)\ \to\ \frac{1}{N} \sum_k \text{Tr}\ g(k),
\end{equation}
with $g(k)$ the quantum metric.
For a finite lattice, we adopt the replacement
$
    \sum_k \leftrightarrow V\int \frac{\mathrm{d}^dk}{(2\pi)^d}
$
with $V=NV_\text{cell}$.
$\Omega_I$ then measures the averaged `distance' of adjacent Bloch states over the Brillouin zone.
Intuitively, $\Omega_I$ is small insofar as the Bloch projector $P_k=\ket{u_k}\bra{u_k}$ is nearly independent of $k$.
Specifically in 2D, $\Omega_I$ is lower-bounded by the Chern number
according to the inequalities Eq.~\eqref{eq:inequalities} above,
\begin{equation}
    \Omega_I \geqslant V_\text{cell} \int \frac{\mathrm{d}^2k}{(2\pi)^2}\ \abs{F_{xy}(k)} \geqslant \frac{V_\text{cell}}{2\pi}\ \abs{C}.
\end{equation}
And we define the associated length scale as
$
    \xi^{\text{2D}}_\text{geo.} \overset{!}{=} \sqrt{\Omega_I} = \sqrt{\frac{1}{N} \sum_k \text{Tr}\ g(k)}
$.
However, although $\Omega_I$ remains geometric in 2D,
$\widetilde{\Omega}$ does not vanish for a nontrivial Chern band under any optimized gauge due to the topological obstruction.
Therefore we have $\min[\Omega]>\Omega_I$ in general.

In 1D, however, the optimal gauge that leads to maximally localized Wannier basis is known as
the parallel transport gauge, which will be introduced in Sec.~\ref{sec:parallel-transport-gauge} with details.
In this gauge, $\widetilde{\Omega}$ is exactly zero,
and hence the minimal Wannier variance $\min[\Omega]=\Omega_I$ is entirely a quantum geometric quantity in 1D.
As a result, we turn to consider the equivalent quasi-1D system described by hybrid Wannier functions localized exponentially in one direction, e.g. the $x$ direction,
and define $\Omega_I(k_y)$ as an analog to $\Omega_I$ in 1D,
\begin{equation}\label{eq:omega-I}
    \Omega_I(k_y) = a \int \frac{\mathrm{d}k_x}{2\pi}\ g_{xx}(k_x,k_y)\ \to\ \frac{1}{L} \sum_{k_x} g_{xx}(k_x,k_y).
\end{equation}
Also we extend $\xi_\text{geo.}$ to quasi 1D,
\begin{equation}
    \xi_\text{geo.}(k_y) \overset{!}{=} \sqrt{\Omega_I(k_y)} = \sqrt{\frac{1}{L} \sum_{k_x} g_{xx}(k_x,k_y)}.
\end{equation}
Since $\min[\Omega(k_y)]=\Omega_I(k_y)$ is fully quantum geometric in quasi 1D,
$\xi_\text{geo.}(k_y)$ exactly characterizes the spatial spread of maximally localized hybrid Wannier function with momentum $k_y$.
This definition of quantum geometric length is adopted throughout our work
due to its precise connection to the hybrid Wannier basis used for transfer matrix calculations.
We also remark that although we have, by convention, made the hybrid Wannier basis localized in $x$ direction
and incorporated $g_{xx}$ in Eq.~\eqref{eq:omega-I},
$\xi_\text{geo.}(k_y)$ remains unaffected by this convention
due to the relation $g_{xx}(k_x,k_y)=g_{yy}(k_y,k_x)$ specific for the $\pi$-flux model.
To see this, it can be first checked that the Hamiltonian Eq.~\eqref{eq:pi-flux-hamil}
exhibits a hidden anti-unitary symmetry,
combining the reflection along the main diagonal $k\to \bar{k}=(k_y,k_x)$
(in the space of first-quantized Hamiltonian), sublattice exchange and complex conjugation,
\begin{equation}\label{eq:symmetry}
    \tau_x H^\ast(\bar{k}) \tau_x = H(k).
\end{equation}
This symmetry is explicitly broken after disorder is added,
and therefore the localization transition still falls into the unitary class.
The combined symmetry Eq.~\eqref{eq:symmetry} guarantees that
the quantum geometric tensor satisfies
$Q^\ast_{\bar{\mu}\bar{\nu}}(\bar{k}) = Q_{\mu\nu}(k)$, where $\bar{\mu}=y$ if $\mu=x$ and $x$ if $\mu=y$.
As a result, the quantum metric obeys
$g_{xx}(k)=g_{yy}(\bar{k})$ and $g_{xy}(k)=g_{xy}(\bar{k})$ as claimed.
Also, the Berry curvature is symmetric as $F_{xy}(k)=F_{xy}(\bar{k})$.

\begin{figure}[htbp]
    \centering\hspace{-1cm}
    \includegraphics[width=.4\columnwidth]{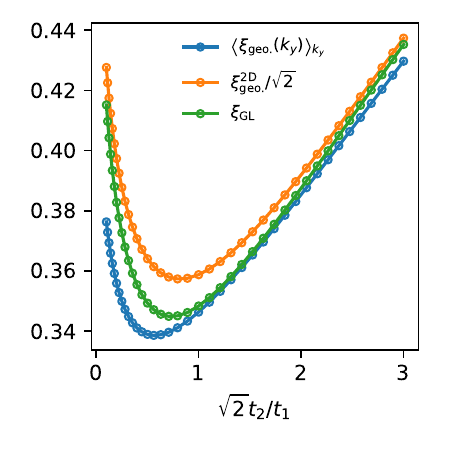}
    \caption{%
        Quantum geometric lengths of the $\pi$-flux model.
    }
    \label{fig:figs3}
\end{figure}

It is notable that in Refs.~\cite{chen2024ginzburglandau,hu2025anomalous},
another topological length
$\xi_\text{GL}=\left\{\det\left[\frac{1}{N}\sum_k g(k)\right]\right\}^{1/4}$
is found to be vital and plays the role of coherence length
in the Ginzburg-Landau theory of flat band superconductors.
In Fig.~\ref{fig:figs3}, we present a comparison of these geometric lengths,
i.e. $\xi^\text{2D}_\text{geo.}$, $\langle\xi_\text{geo.}(k_y)\rangle_{k_y}$, and $\xi_\text{GL}$
for the $\pi$-flux model.
First of all, we note that
\begin{equation}\label{eq:inequality-1}
    \xi^\text{2D}_\text{geo.} \geqslant \sqrt{2}\ \xi_\text{GL}.
\end{equation}
This inequality is saturated when the integrated quantum metric
$\bar{g} = \frac{1}{N}\sum_k g(k) \propto \mathds{1}_{2\times2}$,
because $\text{Tr} g \geqslant 2\sqrt{\det{g}}$ holds for any positive semi-definite $g$.
As suggested before, symmetries shall impose constraints on the quantum metric.
For example, the reflection symmetry along the $x$-axis will eliminate off-diagonal components $\bar{g}_{xy}$.
More importantly, if the model preserves the (combined) $C_4$ symmetry of square lattice,
we have $\bar{g}_{xx}=\bar{g}_{yy}$ and $\bar{g}_{xy}=\bar{g}_{yx}=0$,
such that Eq.~\eqref{eq:inequality-1} must be saturated and $\xi^\text{2D}_\text{geo.}$, $\xi_\text{GL}$ become identical up to a factor of $\sqrt{2}$.
However, except for the combined anti-unitary symmetry Eq.~\eqref{eq:symmetry},
the $\pi$-flux model in the topological phase ($t_1,t_2\neq0$)
breaks these basic, especially $C_4$, symmetries (even when combined symmetries are considered).
Consequently, $\xi^\text{2D}_\text{geo.}$ and $\xi_\text{GL}$ are generally different as shown in Fig.~\ref{fig:figs3}.
In our case, the symmetry in Eq.~\eqref{eq:symmetry} enforces
$\bar{g}_{xx}=\bar{g}_{yy}$, yet in general $\bar{g}_{xy}$ is nonzero.

In addition, it can be proved for the $\pi$-flux model that
$\xi^\text{2D}_\text{geo.}$ and $\langle\xi_\text{geo.}(k_y)\rangle_{k_y}$
satisfy
\begin{equation}\label{eq:inequality-2}
    \xi^\text{2D}_\text{geo.} \geqslant \sqrt{2}\left\langle\xi_\text{geo.}(k_y)\right\rangle_{k_y},
\end{equation}
where $\langle\cdots\rangle_{k_y}$ denotes the average over $k_y$,
and the equality holds if and only if $\Omega_I(k_y)$ is independent of $k_y$.
This follows from $g_{xx}(k_x,k_y)=g_{yy}(k_y,k_x)$ together with Jensen's inequality for the concave function $y=\sqrt{x}$.
The relations in Eqs.~\eqref{eq:inequality-1}\eqref{eq:inequality-2} are then clearly verified in Fig.~\ref{fig:figs3}.
However, the relation between $\langle\xi_\text{geo.}(k_y)\rangle_{k_y}$ and $\xi_\text{GL}$ is unknown in general,
although $\xi_\text{GL}$ seems to be always larger than $\langle\xi_\text{geo.}(k_y)\rangle_{k_y}$.
Moreover, all these quantum geometric lengths share the same order of magnitude,
and exhibit a consistent trend as $t_2/t_1$ varies.
It is therefore challenging to distinguish them when compared with the localization length.

Specifically, for the lowest Landau level with magnetic length $l_B=\sqrt{\frac{1}{\abs{B}}}$,
we note that these quantum geometric length scales coincide as
$
    \xi_\text{geo.}(k_y)=\xi_\text{GL}=\frac{1}{\sqrt{2}}\xi^{\text{2D}}_\text{geo.}=\frac{1}{\sqrt{2}}l_B
$.

\section{Maximally localized hybrid Wannier basis}
The Wannier functions $\ket{W_{nR}}$ form a complete set of localized basis
and are related to the Bloch wavefunction $\ket{\psi_{nk}}$ through a Fourier transformation,
\begin{equation}
    \ket{W_{nR}}_{\alpha,r} = \frac{1}{\sqrt{N}} \sum_k e^{-ikR} \ket{\psi_{nk}}_{\alpha,r},
\end{equation}
where $n$, $\alpha$ are indices for the band and orbital.
According to the Bloch theorem,
$\ket{\psi_{nk}}$ is related to the cell-periodic part $\ket{u_{nk}}$ as
$
    \ket{\psi_{nk}}_{\alpha,r} = e^{ikr} \ket{u_{nk}}_\alpha
$.
It has long been realized that the localization property of Wannier functions depends on the gauge choice of $\ket{u_{nk}}$.
The seminal work of Ref.~\cite{marzari1997maximally} addressed the gauge fixing
and developed the general theory of maximally localized Wannier function in arbitrary dimensions.
In this section, we review the construction of maximally localized (hybrid) Wannier basis
for an isolated band in quasi-1D.

\subsection{Parallel transport gauge}
\label{sec:parallel-transport-gauge}
In 1D, the optimal gauge is known as the \textit{parallel transport gauge}, which yields uniform Berry connections.
The associated maximally localized Wannier functions are eigenstates of the projected position operator $PxP$,
or $Pe^{i\frac{2\pi}{L}x}P$ for periodic systems, with $P$ the projection to occupied bands.
Starting from the cell-periodic states $\ket{u_k}$ with arbitrary gauge,
one can construct the parallel transport gauge $\phi_k$ as follows~\cite{marzari1997maximally,
qi2011generic,marzari2012maximally,lee2013pseudopotential},
\begin{subequations}\begin{align}
    &\qquad\qquad\quad \ket{u_k} \to \ket{\tilde{u}_k} = e^{i\phi_k}\ket{u_k},\\[5pt]
    &\phi_k \overset{!}{=} \int_0^k \mathrm{d}k A_k - k \phi_B, \quad \phi_B = \frac{1}{2\pi}\int_0^{2\pi}\mathrm{d}k A_k.
\end{align}\end{subequations}
$A_k=i\expt{u_k}{\nabla_k u_k}$ denotes the Berry connection,
and $\phi_B$ is the gauge-invariant Berry phase divided by $2\pi$.
Since we only consider the isolated band, the band index is safely omitted.
The Berry connection under this gauge is transformed as
\begin{equation}
    A_k \to \tilde{A}_k = A_k - \nabla_k\phi_k = \phi_B,
\end{equation}
which is uniform and independent of momentum $k$.
It can be checked that the Wannier functions $\ket[false]{\tilde{W}_R}$ under parallel transport gauge
are eigenstates of $PxP$ with eigenvalues $R+\phi_B$ and $P=\sum_k \ket{\psi_k}\bra{\psi_k}$.
Also, the Wannier center $\bar{r}=\expt[false]{\tilde{W}_R}{x}{\tilde{W}_R}=R+\phi_B$,
and $\phi_B$ is thus the charge polarization away from the lattice site $R$.
We see that the $2\pi$ periodicity of Berry phase $2\pi\phi_B$ will shift the Wannier center by an integer number of lattice units.
In order to remove this ambiguity,
we restrict the Berry phase to $\left[0,2\pi\right)$ such that $\phi_B\in\left[0,1\right)$.
As stated earlier in Sec.~\ref{sec:geometric-lengths}, under the parallel transport gauge,
the gauge-dependent Wannier variance $\widetilde{\Omega}$ vanishes in 1D,
and hence the Wannier function is maximally localized with $\Omega=\Omega_I$.
Finally, on a finite lattice, the Berry connection and Berry phase are approximated as
\begin{subequations}\begin{align}
    &\Delta k \cdot A_k = - \text{Arg} \left[\expt{u_k}{u_{k+\Delta k}}\right] = - \text{Im}\ln\left[\expt{u_k}{u_{k+\Delta k}}\right],\\[5pt]
    &\quad\ 2\pi\phi_B = \sum_k \Delta k \cdot A_k = - \text{Im}\ln\left[\prod_k\expt{u_k}{u_{k+\Delta k}}\right].
\end{align}\end{subequations}
As a result, the parallel transport gauge is given by
$
    \phi_k = \sum_{0\leqslant k'<k} \Delta k \cdot A_{k'} - k \phi_B.
$

\subsection{Optimal hybrid Wannier basis in quasi 1D}
For a 2D Chern insulator with nonzero Chern number,
Wannier functions that are exponentially localized in both directions can not be constructed in general.
This is known as the \textit{topological obstruction}.
However, we refer interested readers to Ref.~\cite{gunawardana2024optimally},
which shows that Wannier functions with power-law decay in both directions are still possible for 2D Chern insulators.

Instead, we resort to constructing the \textit{hybrid Wannier function},
which is exponentially localized in one direction while extended along the other.
For a Chern band with $\abs{C}=1$,
the hybrid Wannier functions are counterparts of lowest Landau level wavefunctions
and are particularly convenient for transport and localization calculations
in the presence of spatial disorder due to their localized nature.
The maximally localized hybrid Wannier function is defined as
\begin{equation}
    \ket{W_{k_y,x}} = \frac{1}{\sqrt{L}} \sum_{k_x} e^{-ik_xx} e^{i\phi_{k_x,k_y}} \ket{\psi_{k_x,k_y}},
\end{equation}
where we again assume an isolated band and ignore the band index.
The original 2D model is thus viewed as a quasi-1D system labeled by external momentum $k_y$,
and $\phi_{k_x,k_y}$ denotes the aforementioned parallel transport gauge~\cite{marzari1997maximally,
qi2011generic,marzari2012maximally,lee2013pseudopotential} in 1D,
\begin{equation}
    \phi_{k_x,k_y} = \int_0^{k_x} \mathrm{d}k_x A_x - k_x \phi_{B}(k_y), \quad \phi_{B}(k_y) = \frac{1}{2\pi}\int_0^{2\pi}\mathrm{d}{k_x}A_{x},
\end{equation}
with $A_x$ the $k_x$-component of 2D Berry connection.
In $x$ direction, the hybrid Wannier functions are well localized, and the Wannier centers are given by
\begin{equation}
    r_{k_y,x} = x + \phi_{B}(k_y),
\end{equation}
for $x$ in the bulk.
In quasi 1D, the winding number of $\phi_B(k_y)$ as $k_y$ evolves from $0$ to $2\pi$ yields the Chern number,
\begin{equation}
    \phi_B(2\pi) - \phi_B(0)
    = - \frac{1}{2\pi} \oint_{\partial\text{BZ}} \hat{A} \cdot \mathrm{d}\hat{k}
    = - \frac{1}{2\pi} \int\mathrm{d}^2k\ F_{xy} = -C.
\end{equation}
In the first step, the difference of $\phi_B$ can be considered as
a line integral along the `boundary' of Brillouin zone, which is clockwise and gives rise to the minus sign before the Chern number.
This identity implies that
\begin{equation}
    r_{2\pi,x} - r_{0,x} = -C.
\end{equation}
Namely, the Wannier center shifts by $-C$ as $k_y$ varies from $0$ to $2\pi$.
As revealed in Ref.~\cite{qi2011generic}, for a $\abs{C}=1$ Chern model,
the hybrid Wannier function can be labeled by a single continuous real parameter
$K_y=2\pi x-\text{sgn}\left(C\right)k_y$ with $k_y\in[0,2\pi)$,
which resembles the lowest Landau level wavefunction in the Landau gauge.

\section{Transfer matrix method}
In this section, we develop the transfer matrix method in the maximally localized hybrid Wannier basis.
The $\pi$-flux model on the square lattice is regarded as a quasi-1D system
with the length $L$ along $x$ much larger than the width $M$ along $y$, i.e. $L\gg M$.
Also, we assume the periodic boundary condition (PBC) throughout the discussion.

Firstly, we introduce the onsite disorder operator
\begin{equation}\label{eq:disorder-operator}
    V = \sum_{r,\alpha} V_{r,\alpha} \ket{r,\alpha}\bra{r,\alpha},
\end{equation}
where $V_{r,\alpha}$ are either uniform or Gaussian random variables.
$r$ enumerates spatial sites and $\alpha$ denotes the sublattice index.
Consider the flat-band Hamiltonian projected to the lower-band Wannier subspace,
\begin{equation}\begin{aligned}
    H_\text{flat}
    &= \sum_{xx' k_yk_{y'}} \ket[false]{W_{x,k_y}} \expt[false]{W_{x,k_y}}{V}{W_{x',k_{y'}}} \bra[false]{W_{x',k_{y'}}} \\
    &\overset{!}{=} \sum_{xx' k_yk_{y'}} \left[V_{x,x'}\right]_{y,y'} \ket[false]{W_{x,k_y}} \bra[false]{W_{x',k_{y'}}},
\end{aligned}\end{equation}
where the band energy is set to zero and $V_{x,x'}$ is a $M\times M$ matrix.
Since the Wannier states are localized exponentially in the $x$ direction,
we expect that the entries of $V_{x,x'}$ decay exponentially as $\abs{x'-x}$ increases.
Therefore, we can truncate $V_{x,x'}$ beyond certain length scale $l_0$ and set $V_{x,x'}=0$ for $\abs{x'-x}>l_0$.
The finite cutoff $l_0$ serves as the primary approximation in the transfer matrix method.

Within the isolated-band subspace, the eigenstate $\ket{\psi}$ with eigenenergy $E$ is expanded as
\begin{equation}
    \ket{\psi} = \sum_{x,k_y} a_{x,k_y} \ket{W_{x,k_y}}.
\end{equation}
We define
\begin{equation}
    A_i = \left(a_{i,0}, a_{i,\frac{2\pi}{M}}, \cdots, a_{i,\frac{2\pi(M-1)}{M}}\right)^T,
\end{equation}
for $0\leqslant i<L$, and each $A_i$ is a vector of size $M$.
Under PBC, we identify $\ket[false]{W_{x,k_y}}=\ket[false]{W_{x+nL,k_y}}$ and $A_i=A_{i+nL}$ for $n\in\mathbb{Z}$.
Then the Schr\"odinger equation of $H_\text{flat}$ yields
\begin{equation}\label{eq:schrodinger}
    \sum_{i' j'} \left[V_{i,i'}\right]_{j,j'} \left[A_{i'}\right]_{j'} = E \left[A_i\right]_j
    \quad \Rightarrow \quad
    \sum_{i'} V_{i,i'} A_{i'} = E A_i,
\end{equation}
with $V_{i,i'}$ the $M\times M$ disorder matrix defined beforehand.
Given the truncation $l_0$, Eq.~\eqref{eq:schrodinger} is simplified to
\begin{equation}
    \sum_{|l|\leqslant l_0} V_{i,i+l} A_{i+l} = E A_i
    \quad \Rightarrow \quad
    \sum_{|l|\leqslant l_0} \left(V_{i,i+l} - \delta_{l,0}E\mathds{1}\right) A_{i+l} = 0.
\end{equation}
We then set up the transfer matrix $T^{(i)}$ by first defining the block vectors
\begin{equation}
    \psi^{(i)} = 
    \begin{pmatrix}
        A_{i+l_0-1}\\
        A_{i+l_0-2}\\
        \vdots\\
        A_{i-l_0+1}\\
        A_{i-l_0}
    \end{pmatrix}
    , \quad 0\leqslant i<L,
\end{equation}
which are column vectors with the size of $2l_0M$ and satisfy
$
    \psi^{(i+1)} = \mathbb{T}^{(i)} \psi^{(i)}
$.
The transfer matrix $\mathbb{T}^{(i)}$ has dimension $2l_0M\times 2l_0M$,
\begin{equation}
    \mathbb{T}^{(i)} = 
    \begin{pmatrix}
        -\left[V^i_{l_0}\right]^{-1} V^i_{l_0-1} &
        \cdots &
        -\left[V^i_{l_0}\right]^{-1} \left(V^i_0-E\mathds{1}\right) &
        \cdots &
        -\left[V^i_{l_0}\right]^{-1} V^i_{-l_0+1} &
        -\left[V^i_{l_0}\right]^{-1} V^i_{-l_0} \\[10pt]
        \mathds{1} & \cdots & 0          & \cdots & 0          & 0      \\[10pt]
        \vdots     & \ddots & \vdots     & \ddots & \vdots     & \vdots \\[10pt]
        0          & \cdots & \mathds{1} & \cdots & 0          & 0      \\[10pt]
        \vdots     & \ddots & \vdots     & \ddots & \vdots     & \vdots \\[10pt]
        0          & \cdots & 0          & \cdots & \mathds{1} & 0
    \end{pmatrix},\\[5pt]
\end{equation}
with $0\leqslant i<L$.
For compactness, we have relabeled $V_{i,i+l}$ to $V^{i}_{l}$.
The transfer matrix therefore characterizes the propagation of wavefunction elements along $x$.
Typically, the \textit{localization length} is determined by
calculating the \textit{Lyapunov exponent} of the disordered transfer matrix.
We define
\begin{equation}
    \mathbb{T} = \prod_{i=0}^{L-1} \mathbb{T}^{(i)}, \quad \text{and} \quad \Omega = \frac{1}{2L} \ln \left[\mathbb{T}^\dag \mathbb{T}\right].
\end{equation}
The eigenvalues $\gamma_j$ of the Hermitian matrix $\Omega$ are denoted as the Lyapunov exponents.
In the limit $L\to\infty$, $\gamma_j$ are definite quantities instead of statistical variables
and are expected to come in positive-negative pairs due to the conservation of probability flux.
The localization length $\lambda_M$ is then given by
\textit{the inverse of the smallest positive Lyapunov exponent}~\cite{mackinnon1983scaling,kramer1993localization},
\begin{equation}
    \frac{1}{\lambda_M} = \lim_{L\to\infty} \min_{\gamma_j>0}\left(\gamma_j\right).
\end{equation}
The smallest positive Lyapunov exponent captures the mode with the slowest growth/decay rate of wavefunctions,
whose inverse therefore corresponds to the physical localization length.

\subsection{Numerical stability}
To practically compute $\lambda_M$,
additional efforts should be made to overcome the floating-point overflow and loss of numerical stability during consecutive multiplications of $\mathbb{T}^{(i)}$.
The stable computation is achieved by considering the iterative QR stabilization~\cite{slevin2014critical},
\begin{equation}
    \mathbb{T}^{(i)} \mathbb{Q}^{(i)} = \mathbb{Q}^{(i+1)} \mathbb{R}^{(i+1)},
\end{equation}
where $\mathbb{Q}^{(i)}$ are unitary matrices and $\mathbb{R}^{(i)}$ upper triangular matrices.
In principle, $\mathbb{Q}^{(0)}$ can be chosen arbitrarily as long as it is unitary.
As a result, the accumulated $\mathbb{T}$ satisfies
\begin{equation}
    \mathbb{T} \mathbb{Q}^{(0)} = \mathbb{Q}^{(L)} \left(\prod_{i=1}^{L} \mathbb{R}^{(i)} \right),
\end{equation}
and the localization length is equivalently expressed as
\begin{equation}\label{eq:localization-length}
    \frac{1}{\lambda_M} = \lim_{L\to\infty} \frac{1}{L} \sum_{i=1}^{L} \ln\left(\left\lvert\left[\mathbb{R}^{(i)}\right]_{l_0M-1,l_0M-1}\right\rvert\right).
\end{equation}
This definition of localization length is slightly different from
the aforementioned one related to the eigenvalue of $\Omega$.
However, both definitions converge to the same asymptotic Lyapunov exponent as $L\to\infty$~\cite{slevin2004fluctuations,zhang2005statistics},
guaranteed by the multiplicative ergodic theorem of Oseledec~\cite{crisanti1993products}.
In practice, it is sufficient to perform QR stabilization every few iterations
as long as the condition number of $\mathbb{R}^{(i)}$ does not exceed the numerical precision.

Note that in Eq.~\eqref{eq:localization-length},
we collect only the $l_0M$-th diagonal element of $\mathbb{R}^{(i)}$.
Such simplification is justified by the structure of successive QR factorization,
which progressively orders the orthonormal basis by the asymptotic amplification rates of the corresponding directions in the long-term transfer matrix product.
Therefore, the Lyapunov spectrum,
i.e. the averaged $\overline{\ln\abs{[\mathbb{R}^{(i)}]_{j,j}}}$ for $0\leqslant j< 2l_0M$ as $L\to\infty$,
will be automatically sorted in descending order and symmetric about zero.
(However, due to random fluctuations, individual $\ln\vert[\mathbb{R}^{(i)}]_{j,j}\vert$ at given $i$
are generally not in a strictly descending order with respect to $j$.)
As a result, in order to compute the localization length associated with the smallest positive Lyapunov exponent,
only the $l_0M$-th diagonal element of $\mathbb{R}^{(i)}$ is required.
Moreover, if only the largest $\tilde{M}$ Lyapunov exponents are needed,
we can retain just the first $\tilde{M}$ orthonormal columns of $\mathbb{Q}^{(0)}$ and discard the others.
The dimensions of $\mathbb{Q}^{(i)}$ and $\mathbb{R}^{(i)}$
are then reduced from $2l_0M\times 2l_0M$ to $2l_0M\times \tilde{M}$ and $\tilde{M}\times \tilde{M}$ respectively.
This offers a more efficient and robust calculation of the localization length.
In our case, $\tilde{M}=l_0M$ is used for optimal performance.

\subsection{Computational complexity}
Overall speaking, for a given disorder configuration in Eq.~\eqref{eq:disorder-operator},
the algorithm involves first constructing the $V^i_l$ matrices by evaluating the expectation values,
and performing the standard transfer matrix calculation.
The localization lengths are subsequently averaged over multiple disorder configurations.
To compute each $V^i_l$ with dimension $M\times M$, the complexity involved is $O(LM^3)$
since each element is obtained by a scalar product of the Wannier basis which is of size $n_\text{orb}LM$.
There are in total $\sim l_0L$ number of $V^i_{l}$ to construct during iterations over $x$,
such that the overall complexity reaches $O(l_0 L^2 M^3)$.
On the other hand, the transfer matrix calculations are accomplished with
consecutive multiplications of $V^i_l$ and regular QR stabilization,
whose complexity is $O(l_0 L M^3)$ and $O(l_0^3LM^3)$ respectively.
Consequently, for the considered cylinder geometry $L\gg M,l_0$,
the most computationally expensive part of the algorithm lies in the construction of $V^i_l$,
whose complexity scales as $O(l_0 L^2 M^3)$.

However, noting that the Wannier function is exponentially localized in the $x$ direction,
one can utilize the sparseness of Wannier basis to accelerate the computations of $V^i_l$.
Since the Wannier function decays exponentially,
its amplitudes rapidly fall below the floating-point precision as one moves away from the Wannier center.
Therefore, for sufficiently large $L$, the number of numerically nonzero elements in the Wannier basis is no more than $L_0$,
where $L_0$ does not scale with large $L$ and depends on the localization property of Wannier function.
This sparseness helps to reduce the complexity of the algorithm from $O(l_0 L^2 M^3)$ to approximately $O(l_0 L_0 L M^3)$,
which is now comparable with that of the transfer matrix calculation.

Moreover, for each multiplication of the transfer matrix,
one needs to construct $(2l_0+1)$ number of $V^i_l$ matrices.
And there are in total $(2l_0+1)L$ number of $V^i_l$ matrices to construct
as we sweep over the entire lattice for a fixed disorder configuration.
It is crucial to note that $V^{i+l}_{-l}=\left[V^i_l\right]^\dag$,
and hence $V^i_l$ evaluated at the spatial slice $i$ with $1\leqslant l\leqslant l_0$
can be reused later at $i+l$ as we scan from $i=0$ to $i=L-1$.
Consequently, the number of $V$ matrices evaluated from scratch at each slice is reduced from $2l_0+1$ to $l_0+1$.
The cost of such optimization is that $l_0(l_0+1)/2$ number of cached $V$ matrices with dimension $M\times M$
should be stored in the memory.
Since the calculations of $V$ matrices consume significant computational time in the overall algorithm,
it is highly beneficial that the computational effort of this step is reduced to a fraction of $(l_0+1)/(2l_0+1)$.

\subsection{Convergence with respect to $l_0$}
\begin{figure}[htbp]
    \centering\hspace{-2.25cm}
    \includegraphics[width=.425\columnwidth]{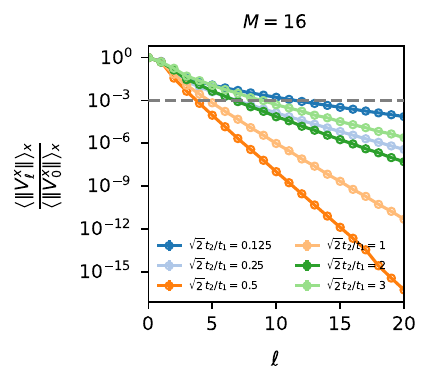}
    \caption{%
        Exponential decay of $\left\Vert V^x_l\right\Vert$ with varied $t_2/t_1$.
    }
    \label{fig:figs4}
\end{figure}

We show in this section the convergence of our calculations with respect to the finite cutoff $l_0$.
As stated earlier, the amplitudes of $V^x_l$ decay exponentially with increasing separation $l$ along $x$.
Practically, we choose $l_0$ by examining the Frobenius norm of $V^x_l$,
\begin{equation}
    l_0=\min\left(l\right), \quad \text{s.t.} \quad \left\langle\left\Vert V^x_l\right\Vert\right\rangle_x / \left\langle\left\Vert V^x_0\right\Vert\right\rangle_x \lesssim 10^{-3},
\end{equation}
where $\langle\cdots\rangle_x$ denotes the average over sites in $x$.
Fig.~\ref{fig:figs4} shows the decay of $\Vert V^x_l\Vert$ for varied $t_2/t_1$.
The characteristic decay length is intrinsic to the localization property of maximally localized Wannier basis,
and is therefore related to the quantum geometric length.
In Table.~\ref{tab:tabs1}, we list the critical exponent $\nu$ extracted with different $l_0$,
and the results converge within the errorbars.

\begin{table}
    \begin{minipage}[htbp]{1\columnwidth}
    \begin{ruledtabular}
    \renewcommand{\arraystretch}{1.1}
    \begin{tabular}{
        c
        c
        c
        c
        c
        S[table-format=1.2(2)]
        c
        S[table-format=1.2(2)]
        S[table-format=1.2]
    }
        & & & & \multicolumn{2}{c}{MFS} & \multicolumn{3}{c}{Data Collapse} \\
        \cmidrule(lr){5-6} \cmidrule(lr){7-9}

        {$\sqrt{2}t_2/t_1$} &
        {$l_0$} &
        {$\langle\Vert V^x_{l_0}\Vert\rangle_x/\langle\Vert V^x_0\Vert\rangle_x$} &
        {$M$} &
        {$E/W$} & {$\nu$} &
        {$E/W$} & {$\nu$} & {$\widetilde{\chi}^2$} \\
        \midrule

        \multirow{3}{*}{0.5}
        & 3 & $4\times10^{-3}$ & $[16,64]$ & $[0,0.2]$ & 2.154(15) & $[0,0.2]$ & 2.17(7)  & 1.39 \\
        & 4 & $6\times10^{-4}$ & $[16,64]$ & $[0,0.2]$ & 2.120(14) & $[0,0.2]$ & 2.14(5)  & 0.60 \\
        & 5 & $9\times10^{-5}$ & $[16,64]$ & $[0,0.2]$ & 2.120(14) & $[0,0.2]$ & 2.14(6)  & 0.92 \\
        \midrule

        \multirow{3}{*}{2.0}
        & 4 & $1\times10^{-2}$ & $[16,64]$ & $[0,0.2]$ & 2.453(32) & $[0,0.3]$ & 2.49(9)  & 1.19 \\
        & 5 & $4\times10^{-3}$ & $[16,64]$ & $[0,0.2]$ & 2.442(26) & $[0,0.3]$ & 2.47(12) & 2.20 \\
        & 7 & $8\times10^{-4}$ & $[16,64]$ & $[0,0.2]$ & 2.505(30) & $[0,0.3]$ & 2.51(8)  & 0.97 \\
    \end{tabular}
    \end{ruledtabular}
    \end{minipage}
    \caption{%
        Convergence of the critical exponent $\nu$ with respect to $l_0$. White-Noise disorder is considered.
    }
    \label{tab:tabs1}
\end{table}

\subsection{Finite-size scaling analysis}
\begin{figure}[htbp]
    \centering\hspace{-1.5cm}
    \includegraphics[width=.7\columnwidth]{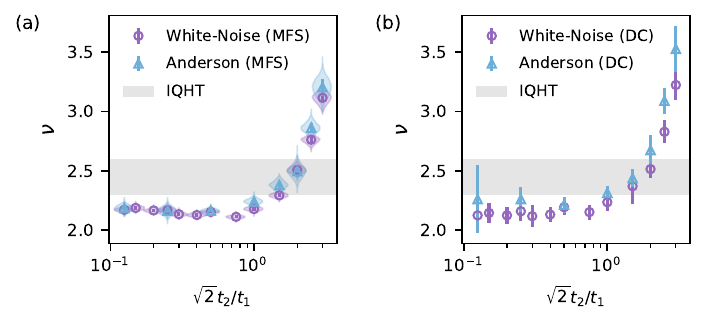}
    \caption{%
        Critical exponent $\nu$ extracted with (a) minimal fitting scheme and (b) data collapse.
        Complementary data is presented in Table.~\ref{tab:tabs2}.
    }
    \label{fig:figs5}
\end{figure}

\begin{figure}[htbp]
    \centering
    \includegraphics[width=1\columnwidth]{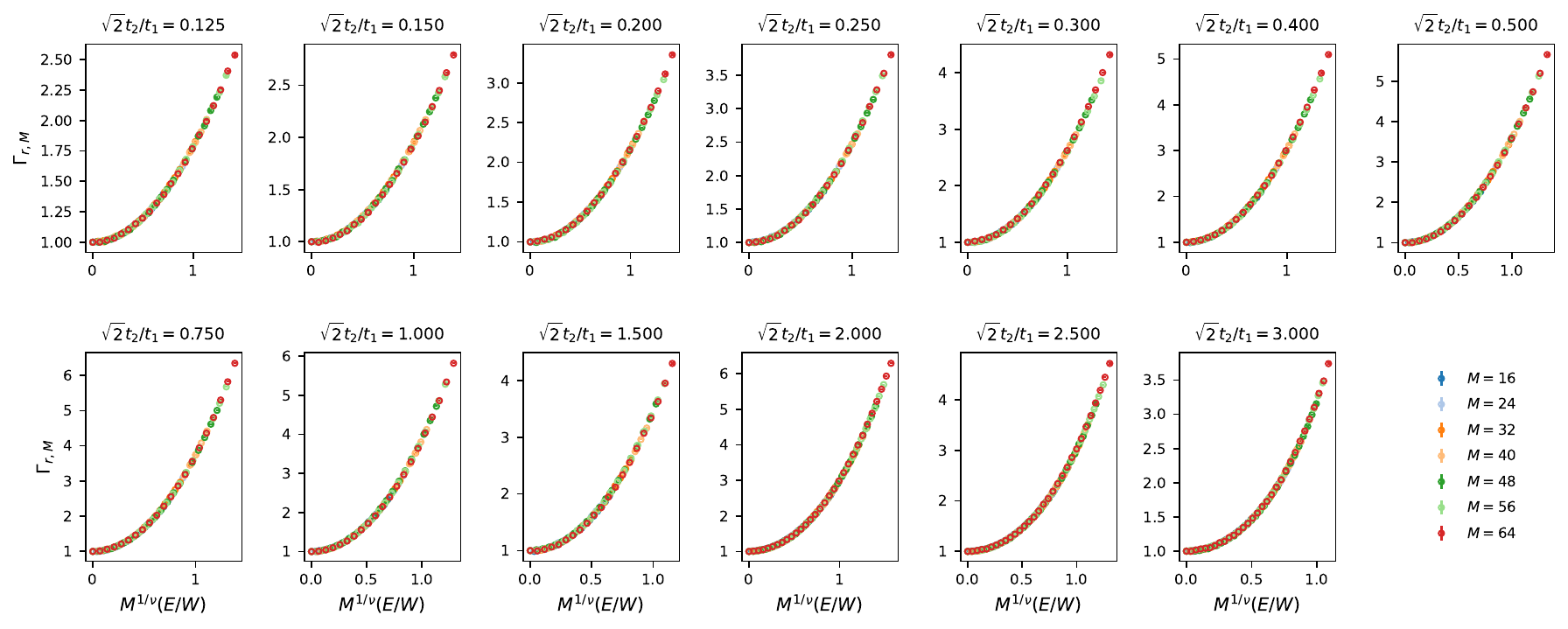}
    \caption{%
        Data collapse of $\Gamma_{r,M}$ for varied $t_2/t_1$ with white-noise disorder.
    }
    \label{fig:figs6}
\end{figure}

\begin{figure}[htbp]
    \centering
    \includegraphics[width=1\columnwidth]{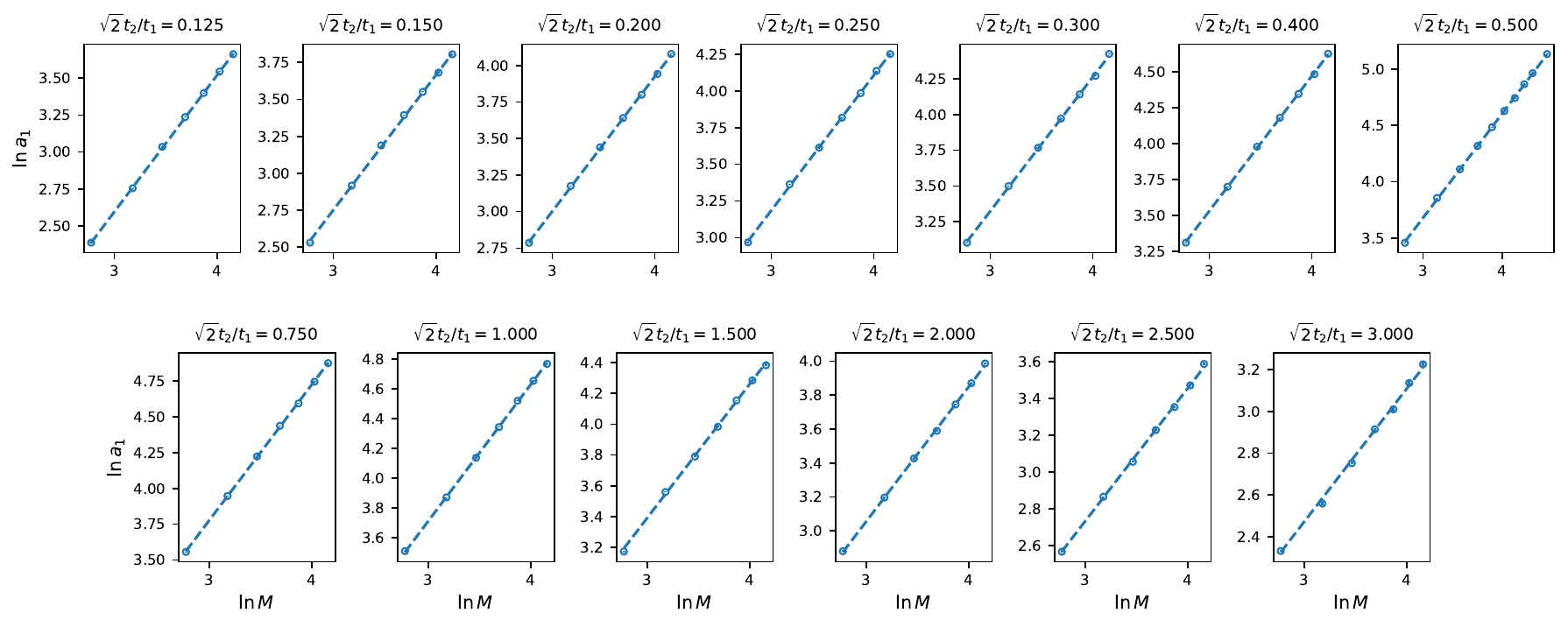}
    \caption{%
        Critical exponent fitting through $\ln a_1\sim \frac{2}{\nu} \ln M$
        in the minimal fitting scheme with white-noise disorder.
    }
    \label{fig:figs7}
\end{figure}

Extensive finite-size scaling data are provided in this section.
To determine the critical exponent, we employ, under the factorization ansatz,
both the minimal fitting scheme discussed in the main text and the data collapse.
The data collapse procedure is conducted with the pyfssa~\cite{sorge2015pyfssa} package,
where a quality function is numerically minimized using a master curve fitted from the data itself,
yielding estimates of the critical exponent $\nu$, its standard deviation, and the reduced $\widetilde{\chi}^2$.
The resulting critical exponents are summarized in Fig.~\ref{fig:figs5} and Table.~\ref{tab:tabs2}.
Fig.~\ref{fig:figs6} demonstrates the quality of data collapse
and Fig.~\ref{fig:figs7} shows the critical exponent fittings in the minimal fitting scheme for varied $t_2/t_1$.

\begin{figure}[htbp]
    \centering
    \includegraphics[width=1\columnwidth]{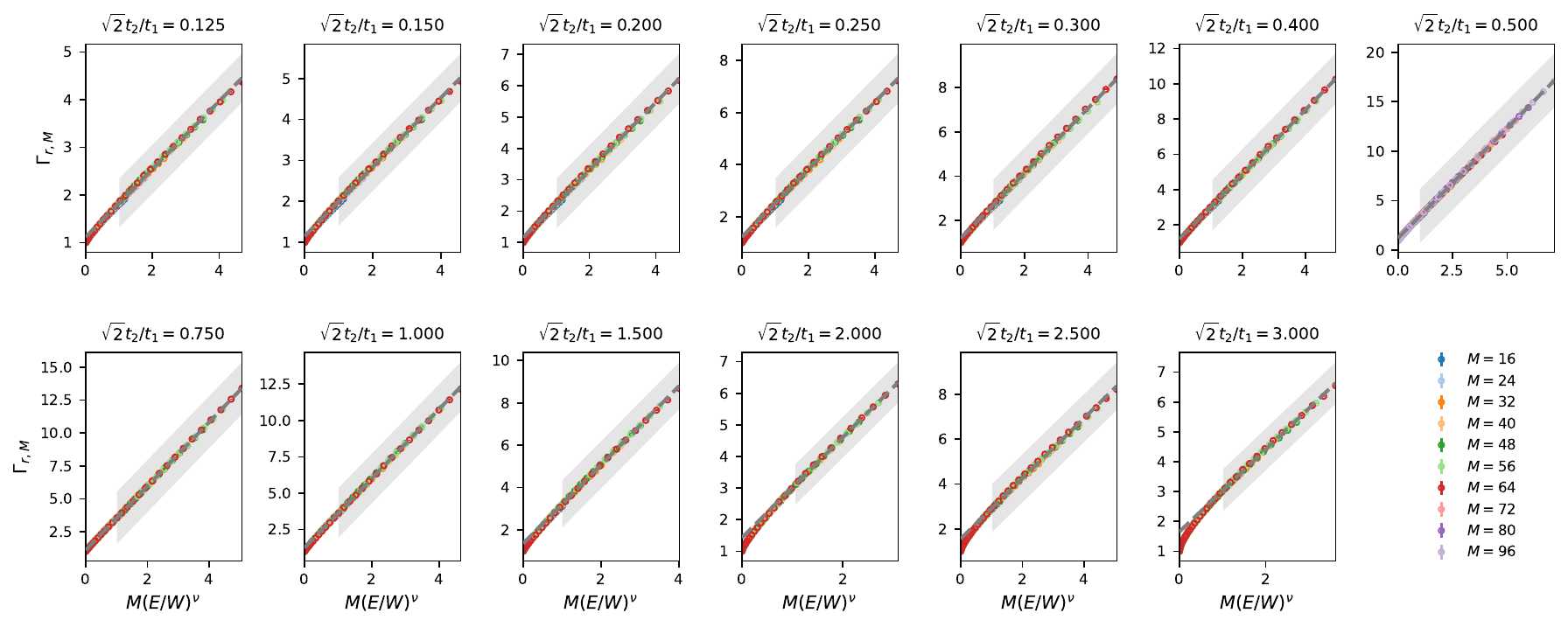}
    \caption{%
        Extraction of $\xi_0$ for varied $t_2/t_1$ with white-noise disorder.
        The grey ribbons indicate the range of data used for the linear fitting of collapsed $\Gamma_r$.
    }
    \label{fig:figs8}
\end{figure}

To finally extract the localization length $\xi_0$,
we plot in Fig.~\ref{fig:figs8} the collapsed $\Gamma_{r,M}$ as a function of $z\overset{!}{=}M(E/W)^\nu=M\xi_0/\xi$,
with $\nu$ the critical exponent determined above.
As indicated by the grey ribbons in Fig.~\ref{fig:figs8},
the collapsed curve $\Gamma_r(z)$ at large $z$ is asymptotically linear in $z$ with slope
$
    s = \frac{1}{\xi_0 \lim_{M\to\infty}\Gamma_M(0)}
$.
We therefore fit the slope $s$ and estimate
$
    \xi_0 = \frac{1}{s\max_M[\Gamma_M(0)]}
$
using the largest accessible $M$.

\begin{table}
    \begin{minipage}[htbp]{1\columnwidth}
    \begin{ruledtabular}
    \renewcommand{\arraystretch}{1.1}
    \begin{tabular}{
        S[table-format=1.3]
        c
        c
        c
        S[table-format=1.2(2)]
        c
        S[table-format=1.2(2)]
        S[table-format=1.2]
    }
        & & & \multicolumn{2}{c}{MFS} & \multicolumn{3}{c}{Data Collapse} \\
        \cmidrule(lr){4-5} \cmidrule(lr){6-8} 
        
        {$\sqrt{2}t_2/t_1$} & {$l_0$} & {$M$} & 
        {$E/W$} & {$\nu$} & 
        {$E/W$} & {$\nu$} & {$\widetilde{\chi}^2$} \\
        \midrule

        \multicolumn{8}{c}{\textit{White-Noise Disorder}} \\
        \midrule

        0.125 & 10 & $[16,64]$ & $[0,0.2]$ & 2.172(17) & $[0,0.2]$ & 2.12(8)  & 0.79 \\
        0.15  & 10 & $[16,64]$ & $[0,0.2]$ & 2.187(16) & $[0,0.2]$ & 2.15(8)  & 1.40 \\
        0.2   & 8  & $[16,64]$ & $[0,0.2]$ & 2.166(15) & $[0,0.2]$ & 2.12(7)  & 1.22 \\
        0.25  & 7  & $[16,64]$ & $[0,0.2]$ & 2.168(19) & $[0,0.2]$ & 2.16(8)  & 0.90 \\
        0.3   & 6  & $[16,64]$ & $[0,0.2]$ & 2.134(16) & $[0,0.2]$ & 2.12(9)  & 1.73 \\
        0.4   & 5  & $[16,64]$ & $[0,0.2]$ & 2.127(14) & $[0,0.2]$ & 2.13(6)  & 1.19 \\
        0.5   & 4  & $[16,96]$ & $[0,0.2]$ & 2.153(11) & $[0,0.2]$ & 2.19(7)  & 1.16 \\
        0.75  & 4  & $[16,64]$ & $[0,0.2]$ & 2.112(13) & $[0,0.2]$ & 2.15(6)  & 0.98 \\
        1.0   & 5  & $[16,64]$ & $[0,0.2]$ & 2.178(16) & $[0,0.2]$ & 2.23(7)  & 1.11 \\
        1.5   & 6  & $[16,64]$ & $[0,0.2]$ & 2.293(18) & $[0,0.2]$ & 2.37(15) & 1.91 \\
        2.0   & 7  & $[16,64]$ & $[0,0.2]$ & 2.507(29) & $[0,0.3]$ & 2.51(8)  & 0.97 \\
        2.5   & 8  & $[16,64]$ & $[0,0.2]$ & 2.762(31) & $[0,0.3]$ & 2.83(10) & 1.47 \\
        3.0   & 9  & $[16,64]$ & $[0,0.2]$ & 3.115(47) & $[0,0.3]$ & 3.22(13) & 1.59 \\
        \midrule
        
        \multicolumn{8}{c}{\textit{Anderson Disorder}} \\
        \midrule

        0.125 & 10 & $[16,64]$ & $[0,0.2]$ & 2.186(25) & $[0,0.15]$ & 2.26(29) & 1.29 \\
        0.25  & 7  & $[16,64]$ & $[0,0.2]$ & 2.164(32) & $[0,0.3]$  & 2.26(11) & 1.16 \\
        0.5   & 4  & $[16,96]$ & $[0,0.2]$ & 2.165(15) & $[0,0.3]$  & 2.21(7)  & 1.21 \\
        1.0   & 5  & $[16,72]$ & $[0,0.2]$ & 2.241(23) & $[0,0.4]$  & 2.31(6)  & 1.16 \\
        1.5   & 6  & $[16,64]$ & $[0,0.2]$ & 2.380(33) & $[0,0.4]$  & 2.44(8)  & 1.13 \\
        2.0   & 7  & $[16,64]$ & $[0,0.2]$ & 2.496(52) & $[0,0.4]$  & 2.67(12) & 1.15 \\
        2.5   & 8  & $[16,64]$ & $[0,0.2]$ & 2.863(44) & $[0,0.4]$  & 3.09(11) & 1.23 \\
        3.0   & 9  & $[16,64]$ & $[0,0.2]$ & 3.205(73) & $[0,0.3]$  & 3.53(20) & 0.94 \\
    \end{tabular}
    \end{ruledtabular}
    \end{minipage}
    \caption{%
        Extensive information on the critical exponent extraction,
        further including the cutoff length $l_0$, range of system sizes, and energy window.
    }
    \label{tab:tabs2}
\end{table}

\end{document}